\documentclass[12pt]{article}
\usepackage{amssymb}
\usepackage{amsmath}
\usepackage{amsfonts}
\usepackage{geometry}
\usepackage{mathrsfs}
\usepackage{epic}
\usepackage{setspace}
\usepackage[title]{appendix}

\newtheorem{theorem}{Theorem}

\newtheorem{corollary}[theorem]{Corollary}
\newtheorem{criterion}[theorem]{Criterion}

\newtheorem{lemma}[theorem]{Lemma}

\newtheorem{proposition}[theorem]{Proposition}

\newenvironment{proof}[1][Proof]{\noindent\textbf{#1.} }{\ \rule{0.5em}{0.5em}}
\input{tcilatex}
\begin{document}

\title{\textbf{Matrix reduction and}\\
\textbf{Lagrangian submodules}}
\author{Olivier Albouy$^{1,2,3}$}
\date{}
\maketitle

$^{1}$ Universit\'{e} de Lyon, F-69622, Lyon, France

$^{2}$ Universit\'{e} Lyon 1, Villeurbanne

$^{3}$ CNRS/IN2P3, UMR5822, Institut de Physique Nucl\'{e}aire de Lyon

\bigskip

E-mail: o.albouy@ipnl.in2p3.fr

\bigskip

\begin{center}
\textbf{Abstract}
\end{center}

This paper deals with three technical ingredients of geometry for quantum
information. Firstly, we give an algorithm to obtain diagonal basis matrices
for submodules of the $%
\mathbb{Z}
_{d}$-module $%
\mathbb{Z}
_{d}^{n}$ and we describe the suitable computational basis. This algorithm
is set along with the mathematical properties and tools that are needed for
symplectic diagonalisation. Secondly, with only symplectic computational
bases allowed, we get an explicit description of the Lagrangian submodules
of $%
\mathbb{Z}
_{d}^{2n}$. Thirdly, we introduce the notion of a fringe of a Gram matrix
and provide an explicit algorithm using it in order to obtain a diagonal
basis matrix with respect to a symplectic computational basis whenever
possible. If it is possible, we call the corresponding submodule nearly
symplectic. We also give an algebraic property in order to single out
symplectic submodules from nearly symplectic ones.

\bigskip



\section*{Introduction}

In recent years, quantum information has grown with increasing interest and
speed. The widest known stimulation for that is the hope of a much more
efficient treatment of information with nanocircuits on the one hand and
quantum algorithms on the other hand. All this is thought to be achievable,
both theoretically and practically, by exploiting the main features of
quantum physics, namely state superposition and entanglement. Let us be more
specific as to quantum algorithms. There have been but a few of them
available till now, but as shown for instance by the Bennett and Brassard's
algorithm BB84~\cite{BB84} and its generalisations for secure communication
or on the contrary the Shor's algorithm~\cite{Shor.94} the realisation of
which would break the widespread RSA code by factorising integers in
polynomial time, they often rely on the discrete Fourier transform (DFT).
This core transformation is only a particular case of another major topic in
quantum theory, namely mutually unbiased bases (MUBs). The characteristic of
such a set of bases is that a state picked out of one of them has equal
amplitude over the states of any other one. In the matrix representing the
discrete Fourier transform, every entry has the same modulus. Thus the basis
one gets by means of the DFT is unbiased with the computational basis.

Besides Fourier transform, the notion of MUBs is widespread both in
classical and quantum information theory. Schwinger unveiled them as soon as
1960 in a paper about unitary operators but he did not name them~\cite%
{Schwinger.60}. They appear in quantum tomography~\cite{WF.89} and in
quantum games such as the Mean King problem~\cite{Vaidman.al.87}\cite%
{Hayashi.al.05}\cite{Paz.al.04}. As to classical information theory, one
finds them in the study of Kerdock codes~\cite{Calderbank.al.97} and
spherical codes~\cite{Renes.04.SphCod} or in the developpement of network
communication protocols~\cite{Renes.04.OptProd}\cite{Heath.al.06}.

Since the beginning of their study in the 80's~\cite{Ivanovic.81}\cite{WF.89}%
, we know that a set of MUBs in a $d$-dimensional Hilbert space contains at
most $d+1$ of them and that this upper-bound can be achieved if $d$ is power
of a prime. But whenever $d$ is a composite integer and despite an extensive
range of mathematics involved, no conclusive information is available about
the achievement of the upper-bound. As a nonexhaustive list of the
mathematical tools that have been used, let us cite Galois fields and rings
in relation with Gaussian sums~\cite{WF.89}\cite{Klapp.Rott.03}\cite%
{Archer.03}, combinatorics, latin squares~\cite{Hayashi.al.05}, unitary
operator bases~\cite{Schwinger.60}\cite{Bandyopadhyay.al.02}, discrete phase
space~\cite{Wootters.04}\cite{Vourdas.05} and Wigner functions~\cite%
{Paz.al.04}\cite{Galvao.05}\cite{Pitt.Rub.05}, Fourier transform~\cite%
{Klimov.al.05.1}\cite{Klimov.al.05.2}, finite ring geometry~\cite{SPR.04}%
\cite{San.Pla.05}\cite{PRPS.06} and also $SU(n)$ Lie groups and their
corresponding Lie algebras~\cite{Kibler.06}\cite{Kib.Pla.06}\cite%
{Alb.Kib.07.1} with connection to positive operator-valued measures (POVMs)~%
\cite{Alb.Kib.07.2}.

Several definitions of the Pauli matrix group have been given throughout
these works. But any of them will be satisfactory for our purpose. Starting
from a paper by Bandyopadhay \textit{et al.}~\cite{Bandyopadhyay.al.02} and
from a study of the Mermin square~\cite{PSK.06}, the strain of finite
geometry has addressed the issue of finitely generated modules over $%
\mathbb{Z}
_{d}$. It appears as a useful, arithmetical translation of the behaviour of
the Pauli operators. The Pauli group divided by its center group is
isomorphic to a $%
\mathbb{Z}
_{d}$-module and by the same token, a tensor product of Pauli groups gives
rise to the direct sum of the corresponding $%
\mathbb{Z}
_{d}$-modules. Despite this isomorphism is related to a quotient group,
commutation relations among the Pauli operators themselves and their ability
to yield MUBs can be translated as geometrical features in the $%
\mathbb{Z}
_{d}$-module we have just mentioned. In particular, the symplectic inner
product, Lagrangian submodules and projective nets appeared to be the
objects of interest. About the connection between MUBs and Lagrangian
submodules, the Heisenberg group and nice error bases, see~\cite{Howe.05}
and \cite{Kibler.08}. The first of those two papers takes place in the frame
of Galois fields. The use of projective lines is illustrated in~\cite%
{Pla.San.08}\cite{Pla.Bab.07} and other references therein, and their study
in relation with their underlying $%
\mathbb{Z}
_{d}$-module is started in~\cite{Hav.San.07}\cite{Hav.San.08}. Moreover, the
action of the Clifford group over a given Pauli group has its own
geometrical counterpart in the $%
\mathbb{Z}
_{d}$-module and can be studied as such.

In this paper, we give a set of tools in order to delve into the structure
of the submodules one meets in quantum theory. Thus in Section~\ref{Simple
reduction section} we deal with basic manipulations of matrices over $%
\mathbb{Z}
_{d}$ and simple diagonalisation. Note that the reduction in question is
that of matrices whose column vectors form a basis of a given submodule, not
of matrices representing linear maps. The properties and tools we introduce
in that section are then used in the frame of symplectic reduction. In
Section~\ref{Symplectic reduction Section}, we build an algorithm in order
to reduce basis matrices to a particular form using only symplectic changes
of basis. This algorithm enables us to set a description of the Lagrangian
submodules of $%
\mathbb{Z}
_{d}^{2n}$ in Section~\ref{Lagrangian submodules}. Finally, the issue of
symplectic diagonalisation is completed for its own sake in Section~\ref{A
criterion for symplectic diagonalisation}.

This paper is primarily intended to physicists and computer scientists
coming to quantum information with various backgrounds. The mathematical
tools involved are all elementary. However, to make the paper
self-contained, we recall every feature of interest for our particular
purpose in two appendices.

\section{Simple reduction\label{Simple reduction section}}

Let $d$ be any integer $\geq 2$. For any specific notations, the reader is
referred to the appendices. As is the case for vector space theory over a
field, vectors in finitely generated modules and linear maps between such
modules can be represented by matrices. The canonical computational basis
for vectors will be denoted $e$. A $k\times l$\ matrix $m$\ is
upper-triangular (resp.~lower-triangular)\ if for all $i\in \{1,\ldots ,k\}$%
, $j\in \{1,\ldots ,l\}$, $i>j$ (resp.~$i<j$), we have $m_{ij}=0$. The
matrix $m$ is diagonal if for all $i\in \{1,\ldots ,k\}$, $j\in \{1,\ldots
,l\}$, $i\neq j$, we have $m_{ij}=0$. The $m_{ii}$'s\ of any matrix will be
called its diagonal coefficients. We extend to matrices the factor
projections $\pi _{p}$ defined in the Chinese remainder theorem (see
Appendix~\ref{CRT}): If $m$ is a $k\times l$ matrix over $%
\mathbb{Z}
/d%
\mathbb{Z}
$ and $p$ is a prime factor of $d$, then $\pi _{p}(m)$ is the $k\times l$\
matrix over $%
\mathbb{Z}
/p^{s}%
\mathbb{Z}
$, $s=v_{p}(d)$,\ whose $(i,j)$ coefficient is $\pi _{p}(m_{ij})$. Also $p$%
-valuation is extended to matrices:%
\begin{equation}
v_{p}(m)=\min (v_{p}(m_{ij});i\in \{1,\ldots ,k\},j\in \{1,\ldots ,l\}).
\end{equation}%
Throughout the paper, we will adopt the conventions that a $\ast $ in a
matrix denotes an arbitrary or unknown coefficient or submatrix, and a blank
denotes a null coefficient or submatrix. The $k\times k$ indentity matrix
will be written $I_{k}$ and the $k\times l$\ null matrix $0_{k,l}$ if
necessary.

In this section and the next one, we address trigonalisation and
diagonalisation of matrices whose columns are basis vectors of a submodule
of $%
\mathbb{Z}
_{d}^{n}$. A left-multiplication by an invertible matrix is to be
interpreted either as an active transformation, that is to say an
automorphism of $%
\mathbb{Z}
_{d}^{n}$, or as a passive transformation, that is to say a change of
computational (free) basis. A right-multiplication by an invertible matrix
stands for a change of basis of the submodule under consideration. The
structure of the given submodule will be much easier to study after
reduction. The reader interested in a more abstract treatment of simple
reduction and in particular diagonalisation of matrices over more general
rings may have a look to \cite{Brown.93}\cite{Kaplansky.49}\cite{LLS.74}. By
the way, we shall also have an insight into generalisation over $%
\mathbb{Z}
_{d}$ of the "Incomplete basis theorem". The set of invertible matrices over 
$%
\mathbb{Z}
$ is denoted $\func{GL}(n,%
\mathbb{Z}
)$ and the set of invertible matrices over $%
\mathbb{Z}
_{d}$ is denoted $\func{GL}(n,%
\mathbb{Z}
_{d})$. Note that left-multiplication by an invertible matrix does not
modify the order of a column vector and hence does not modify the $\gcd $ of
its coefficients. The same is true for right-multiplication and row-vectors.

The only preliminary result we shall admit is that a square matrix with
coefficients in a commutative ring is invertible iff its determinant is an
invertible element of that ring (see \cite{Brown.93}). In fact, the proof is
a mere copy of the field case.

Before we go on, a general remark is in order about the algorithms presented
in this paper. Except the algorithm $\mathscr{D}_{\omega }$ for symplectic
diagonalisation, they are "blind" algorithms, that is to say we do not
suppose we know where invertible coefficients are located in the matrices,
what would be mandatory to use the classical Gaussian reduction for instance.

\begin{lemma}
\label{Reduction of a vector}Let $a\in 
\mathbb{Z}
_{d}^{n}$ be an $n$-dimensional vector. Then%
\begin{equation}
\exists L\in \func{GL}(n,%
\mathbb{Z}
_{d}),\exists k\in 
\mathbb{Z}
_{d},La=ke_{1}.
\end{equation}%
The column vectors $C_{1},\ldots ,C_{n}$\ of $L^{-1}$ form a free basis of $%
\mathbb{Z}
_{d}^{n}$ such that $kC_{1}=a$.
\end{lemma}

\begin{proof}
Our calculations to prove this lemma will be in $%
\mathbb{Z}
$. The results will only have to be sent onto residue classes at the end.
Let $a\in 
\mathbb{Z}
^{n}$, $\delta _{n-1}=a_{n-1}\wedge a_{n}$, $a_{n-1}^{\prime
}=a_{n-1}/\delta $, $a_{n}^{\prime }=a_{n}/\delta $. There exist $%
k_{1},l_{1}\in 
\mathbb{Z}
$ such that $k_{1}a_{n-1}+l_{1}a_{n}=\delta _{n-1}$ so that we have the
active transformation on $a$:%
\begin{equation}
\begin{array}{cccc}
\underbrace{\left( 
\begin{array}{c|cc}
I_{n-2} &  &  \\ \hline
& k_{1} & l_{1} \\ 
& -a_{n}^{\prime } & a_{n-1}^{\prime }%
\end{array}%
\right) } & \underbrace{\left( 
\begin{array}{c}
\ast \\ \hline
a_{n-1} \\ 
a_{n}%
\end{array}%
\right) } & = & \underbrace{\left( 
\begin{array}{c}
\ast \\ \hline
\delta _{n-1} \\ 
0%
\end{array}%
\right) }. \\ 
L^{(n-1)}\in \func{GL}(n,%
\mathbb{Z}
) & a &  & a^{(n-1)}%
\end{array}%
\end{equation}%
Repeating this trick on $a^{(n-1)}$ with components $n-1$ and $n-2$ and so
on, we bring the vector $a$ onto a multiple of $e_{1}$. Of course, the order
of $k$ in $%
\mathbb{Z}
_{d}$ is the same as the order of $a$ in $%
\mathbb{Z}
_{d}^{n}$. In details:

$a^{(n)}=a,\delta _{n}=a_{n},$

$\forall i\in \{1,\ldots ,n-1\},\left\{ 
\begin{array}{l}
\begin{array}{rcl}
\delta _{n-i} & = & a_{n-i}\wedge \delta _{n-i+1} \\ 
a_{n-i}^{\prime } & = & a_{n-i}/\delta _{n-i} \\ 
\delta _{n-i+1}^{\prime } & = & \delta _{n-i+1}/\delta _{n-i}%
\end{array}%
\medskip \\ 
\exists k_{i},l_{i}\in 
\mathbb{Z}
_{d},k_{i}a_{n-i}+l_{i}\delta _{n-i+1}=\delta _{n-i}%
\end{array}%
\right. ,$

\bigskip

\begin{equation}
\begin{array}{cccc}
\underbrace{\left( 
\begin{array}{c|cc|c}
I_{n-i-1} &  &  &  \\ \hline
& k_{i} & l_{i} &  \\ 
& -\delta _{n-i+1}^{\prime } & a_{n-i}^{\prime } &  \\ \hline
&  &  & I_{i-1}%
\end{array}%
\right) } & \underbrace{\left( 
\begin{array}{c}
\ast \\ \hline
a_{n-i} \\ 
\delta _{n-i+1} \\ \hline
0_{i-1,1}%
\end{array}%
\right) } & = & \underbrace{\left( 
\begin{array}{c}
\ast \\ \hline
\delta _{n-i} \\ 
0 \\ \hline
0_{i-1,1}%
\end{array}%
\right) }. \\ 
L^{(n-i)} & a^{(n-i+1)} &  & a^{(n-i)}%
\end{array}%
\end{equation}%
Each $L^{(i)}$ has determinant $1$, so that the complete transformation
given by the product $L=\prod_{i=1}^{n-1}L^{(i)}$ also has and therefore is
an automorphism. So we have shown what we were seeking for:%
\begin{equation}
\exists L\in \func{GL}(n,%
\mathbb{Z}
),\exists k\in 
\mathbb{Z}
,La=ke_{1}.
\end{equation}
\end{proof}

\begin{lemma}
\label{Decreasing orders}Let $a_{1},a_{2}\in 
\mathbb{Z}
_{d}^{n}$ of order $\nu _{1},\nu _{2}$ respectively. There exists a linear
combination $a$ of $a_{1},a_{2}$ of order $\nu _{1}\vee \nu _{2}$. Moreover,
if $d$ is odd, we can build $a$ such that%
\begin{equation}
\left\langle a,a_{1}\right\rangle =\left\langle a,a_{2}\right\rangle
=\left\langle a_{1},a_{2}\right\rangle .
\end{equation}%
If $d$ is even, then in general we can have only%
\begin{equation}
\left\langle a,a_{1}\right\rangle \text{ or\ }\left\langle
a,a_{2}\right\rangle =\left\langle a_{1},a_{2}\right\rangle .
\end{equation}
\end{lemma}

\begin{proof}
If $a_{1}$ or $a_{2}$ is equal to $0$, the lemma is obvious. We now suppose
that they are not and that $d$ is odd. Let $A=(a_{1}|a_{2})$ be the $n\times
2$ matrix whose columns are $a_{1},a_{2}$ and with the help of lemma~\ref%
{Reduction of a vector}, left-multiply $A$ by an invertible matrix $L$ such
that $La_{1}$ has all but its first coefficient equal to $0$. The matrix $L$
is to be interpreted as a change of basis. If $k_{1},\ldots ,k_{n}$ are the
coefficients of the second column of $LA$, let $\delta =k\wedge k_{1}$.
According to lemma~\ref{Invertible coeffs for gcd} of Appendix~\ref{CRT},
there exist $u,v\in U(%
\mathbb{Z}
_{d})$ such that%
\begin{equation}
\delta =uk+vk_{1}.
\end{equation}%
Then we put%
\begin{equation}
(a_{1}^{\prime }|a)=LA\left( 
\begin{array}{cc}
0 & u \\ 
-u^{-1} & v%
\end{array}%
\right) \text{ and }(a|a_{2}^{\prime })=LA\left( 
\begin{array}{cc}
u & 0 \\ 
v & u^{-1}%
\end{array}%
\right) .
\end{equation}%
or%
\begin{equation}
(a_{1}^{\prime }|a)=LA\left( 
\begin{array}{cc}
v^{-1} & u \\ 
0 & v%
\end{array}%
\right) \text{ and }(a|a_{2}^{\prime })=LA\left( 
\begin{array}{cc}
u & -v^{-1} \\ 
v & 0%
\end{array}%
\right) .
\end{equation}%
In any case, $a$ answers the lemma since, with lemma~\ref{Preserving order
in Zd} and equations (\ref{Calculation of order 1}), (\ref{Calculation of
order 2}) and (\ref{Calculation of order 4}) of the appendices, the order of 
$a$ is%
\begin{multline}
\nu (a)=\frac{d}{\delta \wedge \left(
\bigwedge\nolimits_{i=2}^{n}k_{i}\right) \wedge d}=\frac{d}{\left( k\wedge
d\right) \wedge \left( \bigwedge\nolimits_{i=1}^{n}k_{i}\wedge d\right) } \\
=\left( \frac{d}{k\wedge d}\right) \vee \left( \frac{d}{\bigwedge%
\nolimits_{i=1}^{n}k_{i}\wedge d}\right) =\nu (a_{1})\vee \nu (a_{2}).
\end{multline}%
And for $i=1,2$,%
\begin{equation}
\left\langle a,a_{i}\right\rangle =\left\langle a,a_{i}^{\prime
}\right\rangle =\left\langle a_{1},a_{2}\right\rangle .
\end{equation}

To complete the proof, let us deal with the case where $d=2^{s}$. With $i=1$
or $2$ such that $\nu (a_{i})=\min (\nu (a_{1}),\nu (a_{2}))$, we simply put 
$a=a_{i}$.
\end{proof}

Note that for any linear combination $b=b_{1}a_{1}+b_{2}a_{2}$ of $a_{1}$
and $a_{2}$,%
\begin{equation}
\nu (a)b=b_{1}(\nu (a)a_{1})+b_{2}(\nu (a)a_{2})=0.
\end{equation}%
Thus for all $b\in \left\langle a_{1},a_{2}\right\rangle $, $\nu (b)$
divides $\nu (a)$.

\bigskip

Given two minimal bases $f=(f_{1},\ldots ,f_{r})$ and $g=(g_{1},\ldots
,g_{r})$ of a submodule $M$, it is in general not possible to find an
automorphism of $M$ that brings $f_{i}$ onto $g_{i}$ for all $i$, even if $%
\nu (f_{i})=\nu (g_{i})$ for all $i$. Indeed in $%
\mathbb{Z}
_{6}$, we cannot find $a\in 
\mathbb{Z}
_{6}$ and $b\in U(%
\mathbb{Z}
_{6})$ so that%
\begin{equation}
\left( 
\begin{array}{cc}
1 & a \\ 
0 & b%
\end{array}%
\right) \left( 
\begin{array}{cc}
1 & 1 \\ 
0 & 3%
\end{array}%
\right) =\left( 
\begin{array}{cc}
1 & 2 \\ 
0 & 3%
\end{array}%
\right) .
\end{equation}%
We can take $b$ to be $1$ or $5$. But $a$ should be such that $1+3a=2$, what
is impossible. As to diagonalisation, left-multiplication is still not
sufficient, especially because the order of respective column vectors from
one basis to the other is not preserved:%
\begin{equation}
\left( 
\begin{array}{cc}
1 & a \\ 
0 & b%
\end{array}%
\right) \left( 
\begin{array}{cc}
1 & 1 \\ 
0 & 3%
\end{array}%
\right) \neq \left( 
\begin{array}{cc}
1 & 0 \\ 
0 & 3%
\end{array}%
\right) .
\end{equation}%
We shall make use of lemma~\ref{Decreasing orders} to perform
diagonalisation with left- and right-multiplications. For instance, the
latter inequation is solved trivially:%
\begin{equation}
\left( 
\begin{array}{cc}
1 & 0 \\ 
0 & 1%
\end{array}%
\right) \left( 
\begin{array}{cc}
1 & 1 \\ 
0 & 3%
\end{array}%
\right) \left( 
\begin{array}{cc}
1 & -1 \\ 
0 & 1%
\end{array}%
\right) =\left( 
\begin{array}{cc}
1 & 0 \\ 
0 & 3%
\end{array}%
\right) .
\end{equation}

Suppose that we are given a minimal basis $b=(b_{1},\ldots ,b_{r})$\ of a
submodule $M$ of $%
\mathbb{Z}
_{d}^{n}$\ and $B$ is the matrix of size $n\times r$\ whose $i$-th column is 
$b_{i}$. The matrix $B$ is called a basis matrix for $M$. With the help of
lemma~\ref{Reduction of a vector}, we could easily put $B$ in an
upper-triangular form by means of left-multiplications. But we are going to
transform it into a new, diagonal matrix whose column vectors still generate 
$M$. Because of lemma~\ref{Decreasing orders} and associativity of $\func{lcm%
}$, we may suppose that%
\begin{subequations}%
\begin{gather}
\nu (b_{1})=\bigvee_{i=1}^{r}\nu (b_{i}), \\
\forall m\in M,\nu (m)|\nu (b_{1}).
\end{gather}%
\end{subequations}%
An algorithm which set any matrix that way will be called $\mathscr{A}$. It
consists of an appropriate right-multiplication by an invertible matrix. We
left-multiply $B$ by a matrix $L_{1}$\ with determinant $1$ so that $%
L_{1}b_{1}$ has all but its first coefficient equal to $0$. Let $\widetilde{B%
}=L_{1}B$. If one of the coefficients of $\widetilde{B}$ but in the first
column, say $\widetilde{b}_{ij}$, $j\geq 2$, were not a multiple of the
upper-left coefficient $\widetilde{b}_{11}$, then $\nu \left( \widetilde{b}%
_{ij}\right) $ would not be a divisor of $\nu \left( \widetilde{b}%
_{11}\right) =\nu (b_{1})$ and according to relation (\ref{Calculation of
order 3}) of Appendix~\ref{FGM Section}\ and to lemma~\ref{Decreasing orders}
again, there would exist a linear combination of $b_{1}$ and $b_{j}$ of
order greater than $\nu (b_{1})$, what is impossible by assumption. Since we
are only interested in a basis of $M$ we can put all but the first
coefficient of the first row to $0$ and obtain a matrix $B_{1}$. This is
equivalent to a right-multiplication by an appropriate invertible matrix.
Carrying on this process, we obtain a diagonal matrix $B_{r}$\ whose column
vectors still form a minimal basis of $M$. Let us describe the algorithm in
details.

\bigskip

\textbf{Algorihtm} $\mathscr{D}_{0}$: The starting point is the empty matrix 
$D_{0}$ with no lines and no columns, and as an argument a $k\times l\ $%
matrix $B$. Let $B_{0}=B$. Then for $i$ from $0$ to $\mu =\min (k-1,l-1)$,
we go on the following steps:

\begin{enumerate}
\item $R_{i+1}^{(1)}=\left( 
\begin{array}{cc}
I_{i} & 0 \\ 
0 & R^{\prime }%
\end{array}%
\right) $ with $R^{\prime }$ a $(l-i)\times (l-i)$ invertible\ matrix such
that $\mathscr{A}(B_{i})=B_{i}R^{\prime }$.

\item $L_{i+1}=\left( 
\begin{array}{cc}
I_{i} & 0 \\ 
0 & L^{\prime }%
\end{array}%
\right) $ with $L^{\prime }$ an $(k-i)\times (k-i)$,\ determinant-$1$\
matrix given by lemma~\ref{Reduction of a vector} such that $B^{\prime
}=L^{\prime }\mathscr{A}(B_{i})$ has all its first column coefficients but
the first one equal to $0$.

\item $R_{i+1}^{(2)}=\left( 
\begin{array}{cc}
I_{i} & 0 \\ 
0 & R^{\prime \prime }%
\end{array}%
\right) $ with $R^{\prime \prime }$ a $(l-i)\times (l-i)$ invertible\ matrix
such that $B^{\prime }R^{\prime \prime }$ has all its first line
coefficients but the first one equal to $0$.

\item $D_{i+1}=\left( 
\begin{array}{cc}
D_{i} & 0 \\ 
0 & b_{11}^{\prime }%
\end{array}%
\right) $.

\item $B_{i+1}$ is given from $B^{\prime }$ by deleting the first row and
the first column of this latter one.
\end{enumerate}

\noindent The results of the algorithm are the change of basis matrices $L(B)%
\label{L in algorithm D0}=\prod_{i=1}^{\mu +1}L_{\mu +2-i}$,\ $%
R(B)=\prod_{i=1}^{\mu +1}R_{i}^{(1)}R_{i}^{(2)}$ and the $k\times l\ $%
diagonal matrix $\mathscr{D}_{0}(B)$ defined to be 
\begin{equation}
\left( 
\begin{array}{c}
D_{\mu +1} \\ 
0_{k-l,l}%
\end{array}%
\right) \text{ or }\left( 
\begin{array}{cc}
D_{\mu +1} & 0_{k,l-k}%
\end{array}%
\right)
\end{equation}%
whether $k\geq l$ or $k\leq l$ respectively. For all $i,j\in \{1,\ldots ,r\}$%
, $i<j$, we have $(D_{\mu +1})_{ii}|(D_{\mu +1})_{jj}$.\quad $\blacklozenge $

\bigskip

As to the minimal basis $b$, the second case for $\mathscr{D}_{0}(B)$ is
impossible and thus $r\leq n$. The minimality of $b$ also implies that none
of the diagonal coefficients of $D_{\mu +1}=D_{r}$ is $0$. Hence, the column
vectors of $\mathscr{D}_{0}(B)$ still form a minimal basis of $M$.
Additionally, note that if we replace every diagonal entry of $\mathscr{D}%
_{0}(B)$ by $1$, the column vectors of the matrix we obtain form a free
basis $\widehat{b}$ of a free, rank-$r$ submodule $M_{\widehat{b}}$
containing $M$.

The remaining features stated in theorem~\ref{Simple reduction theorem}
below are immediate consequences of the classification of finite,
commutative groups. However, we are to prove them as an illustration of our
topic which is reduction of matrices with coefficients in $%
\mathbb{Z}
_{d}$.

If we start with a nonminimal basis of $M$, say $b^{\prime }$ with $%
r+r^{\prime }$ vectors, $r^{\prime }\geq 1$, the algorithm $\mathscr{D}_{0}$
yields a matrix of the form%
\begin{equation}
\mathscr{D}_{0}(B^{\prime })=\left( 
\begin{array}{cc}
D & 0_{n,r+r^{\prime }-k}%
\end{array}%
\right) ,
\end{equation}%
where $D$ is a diagonal matrix with $k$ columns, all of them nonzero. Since $%
M$ is of rank $r$, we have $k\geq r$. Suppose $k>r$ and let $\widetilde{D}$
be the $(1,\ldots ,n;1,\ldots ,r+1)$ submatrix of $D$. There exists an $%
r\times (r+1)$ matrix $E$\ whose $j$-th column, $j\in \{1,\ldots ,r+1\}$,
contains the components of the $j$-th column vector of $\widetilde{D}$\ with
respect to the free basis $\widehat{b}$. A linear combination of the column
vectors of $\widetilde{D}$ with some factors is null iff the linear
combination of the respective column vectors of $E$ with these same factors
is null. In other words, $\widetilde{D}$ and $E$ have the same kernel as
linear maps. Applying the algorithm $\mathscr{D}_{0}$\ to $E$, we construct
a null linear combination of its column vectors the factors of which are
located in the last column $C$ of $R(E)$. Now, let us choose a prime factor $%
p$ of $d$ such that $t=v_{p}(d_{r+1,r+1})<v_{p}(d)$, so that no diagonal
entry of $\pi _{p}\big(\widetilde{D}\big)$ is null. There exists such a $p$
because $d_{r+1,r+1}\neq 0$. Since $R(E)$ is invertible, at least one of the
factors contained in $\pi _{p}(C)$\ is a unit. But in that case, $\pi _{p}%
\big(\widetilde{D}C\big)$ cannot be null as expected. So $k=r$. Thus we may
add to the algorithm $\mathscr{D}_{0}$ a final step to get the

\bigskip

\textbf{Simple reduction algorithm} $\mathscr{D}$: Let $M$ be a rank-$r$
submodule of $%
\mathbb{Z}
_{d}^{n}$, $b$ a basis of $M$ containing $s\geq r$ vectors\ and $B$ the
corresponding basis matrix. By deleting the last $s-r$ null columns of $%
\mathscr{D}_{0}(B)$, one gets a minimal basis matrix for $M$. The matrix $%
\mathscr{D}(b)=\mathscr{D}(B)$ thus obtained is called the simple reduction
of the basis $b$ or of the basis matrix $B$.\quad $\blacklozenge $

\bigskip

Let $b^{(1)}$ and $b^{(2)}$ be two bases of $M$. In the next three
paragraphs, we are going to work in a single Chinese factor, say with prime
factor $p$, and we are to prove that for every $i\in \{1,\ldots ,r\}$, the $%
i $-th diagonal entries of $\mathscr{D}(b^{(1)})$ and $\mathscr{D}(b^{(2)})$
are associated. In order to make notations lighter, we even suppose that $d$
is a power of a prime, say $p^{s}$. There is a slight difference, since in
the latter case, $r$ may vary with the Chinese factor one chose initially.
The reader may check that such a trick is allowed. Let $%
B^{(a)}=L(b^{(a)})^{-1}\mathscr{D}(b^{(a)})$, $a\in \{1,2\}$, $\widehat{B}$
be the representative matrix of $\widehat{b}$ with respect to the
computational basis and $P^{12}$, $P^{21}$ and $E$ be three $r\times r$\
matrices such that%
\begin{equation}
B^{(1)}P^{12}=B^{(2)},\quad B^{(2)}P^{21}=B^{(1)},\quad \widehat{B}E=B^{(1)}.
\end{equation}%
So we have%
\begin{equation}
\widehat{B}EP^{12}P^{21}=B^{(1)}P^{12}P^{21}=B^{(1)}=\widehat{B}E
\end{equation}%
and then

\begin{equation}
\mathscr{D}(E)P=\mathscr{D}(E),\text{ with }P=R(E)^{-1}P^{12}P^{21}R(E).
\end{equation}%
If some diagonal entry of $\mathscr{D}(E)$ were zero, then the column
vectors of $B^{(1)}R(E)=\widehat{B}L(E)^{-1}\mathscr{D}(E)$ would form a
basis of $M$ with at most $r-1$ elements, what is impossible. So there
exists an $r\times r$ matrix $Q$ such that $P=I_{r}+pQ$. Hence $P$ is
invertible, and so are $P^{12}$ and $P^{21}$. For $a\in \{1,2\}$, consider
the maps%
\begin{equation}
\begin{array}{crcl}
f^{(a)}: & (%
\mathbb{Z}
/p^{s}%
\mathbb{Z}
)^{r} & \longrightarrow & M \\ 
& X & \longmapsto & B^{(a)}X%
\end{array}%
\end{equation}%
where elements of $(%
\mathbb{Z}
/p^{s}%
\mathbb{Z}
)^{r}$ are presented as column vectors and let $n_{i}^{(a)}$, $i\in
\{0,\ldots ,s\}$, be the number of vectors $X$ so that $f^{(a)}(X)$ is of
order $p^{s-i}$. For every $X$ so that $B^{(1)}X$ is of order $p^{s-i}$, the
vector $Y=P^{21}X$ is so that $B^{(2)}Y$ is of order $p^{s-i}$ as well.
Since $P^{21}$ is injective as a linear map, we have $n_{i}^{(2)}\geq
n_{i}^{(1)}$. The converse inequality can be shown the same way and so $%
n_{i}^{(1)}=n_{i}^{(2)}$.

Now let $b$ be any basis of $M$ and $r_{i}$, $i\in \{0,\ldots ,s\}$, be the
number of diagonal entries of $D=\mathscr{D}(b)$ of the form $up^{i}$, $u\in
U(%
\mathbb{Z}
/p^{s}%
\mathbb{Z}
)$. We also define the following two related objects%
\begin{equation}
\forall i\in \{-1,\ldots ,s-1\},\sigma _{i}=\sum_{j=0}^{i}r_{j},
\end{equation}%
and as intervals in $%
\mathbb{N}
$%
\begin{equation}
\forall i\in \{0,\ldots ,s-1\},K_{i}=\{\sigma _{i-1}+1,\ldots ,\sigma _{i}\}.
\label{K Definition}
\end{equation}%
The cardinality of a $K_{i}$ is of course $r_{i}$. We are to prove by
induction on $i$ that the $r_{i}$'s do not depend on the choice of $b$ and
so are properties of $M$. As in the previous paragraph, consider the map%
\begin{equation}
\begin{array}{crcl}
f: & (%
\mathbb{Z}
/p^{s}%
\mathbb{Z}
)^{r} & \longrightarrow & M \\ 
& X & \longmapsto & DX.%
\end{array}%
\end{equation}%
The number of vectors $X$ so that $f(X)$ is of order $p^{s-i}$, $i\in
\{0,\ldots ,s-1\}$, is%
\begin{multline}
n_{i}=\sum_{j=0}^{i}\left\{ \left[ \prod_{k=0}^{j-1}p^{((s-1)-(i-k))r_{k}}%
\right] \times \left[ p^{((s-1)-(i-j-1))r_{j}}-p^{((s-1)-(i-j))r_{j}}\right]
\times \right. \\
\left. \times \left[ \prod_{k=j+1}^{i}p^{((s-1)-(i-k-1))r_{k}}\right]
\right\} \times \prod_{l=i+1}^{s-1}p^{sr_{l}}.
\end{multline}%
Indeed, one can consider the bar graph in figure \ref{Calculation of n_i}\
to see where that latter expression comes from.

\unitlength=1cm

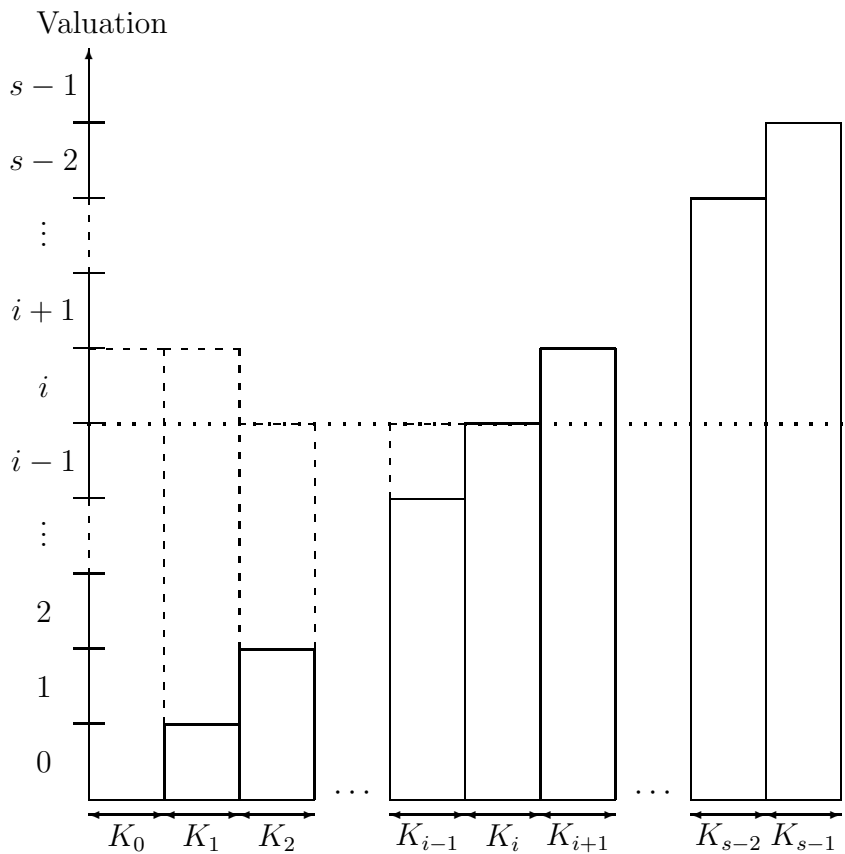
\begin{figure}[th]
\begin{center}
\begin{picture}(12,11.5)(-1,-0.5)

\put(0,0){\line(1,0){3}}
\put(3.25,0){$\cdots$}
\put(4,0){\line(1,0){3}}
\put(7.25,0){$\cdots$}
\put(8,0){\line(1,0){2}}
\multiput(0.5,-0.2)(1,0){3}{\vector(1,0){0.5}}
\multiput(0.5,-0.2)(1,0){3}{\vector(-1,0){0.5}}
\multiput(4.5,-0.2)(1,0){3}{\vector(1,0){0.5}}
\multiput(4.5,-0.2)(1,0){3}{\vector(-1,0){0.5}}
\multiput(8.5,-0.2)(1,0){2}{\vector(1,0){0.5}}
\multiput(8.5,-0.2)(1,0){2}{\vector(-1,0){0.5}}
\multiputlist(0.5,-0.5)(1,0){$K_{0}$,$K_{1}$,$K_{2}$}
\multiputlist(4.5,-0.5)(1,0){$K_{i-1}$,$K_{i}$,$K_{i+1}$}
\multiputlist(8.5,-0.5)(1,0){$K_{s-2}$,$K_{s-1}$}

\put(0,0){\line(0,1){3}}
\dashline{0.1}(0,3)(0,4)
\put(0,4){\line(0,1){3}}
\dashline{0.1}(0,7)(0,8)
\put(0,8){\vector(0,1){2}}
\multiput(-0.2,1)(0,1){9}{\rule{0.4cm}{0.5pt}}
\put(-0.7,10.2){Valuation}
\multiputlist(-0.6,0.5)(0,1){$0$,$1$,$2$,$\raisebox{1.5ex}{\vdots}$,$i-1$,$i$,$i+1$,$\raisebox{1.5ex}{\vdots}$,$s-2$,$s-1$}

\drawline(1,0)(1,1)(2,1)
\drawline(2,0)(2,2)(3,2)(3,0)
\drawline(4,0)(4,4)(5,4)
\drawline(5,0)(5,5)(6,5)
\drawline(6,0)(6,6)(7,6)(7,0)
\drawline(8,0)(8,8)(9,8)
\drawline(9,0)(9,9)(10,9)(10,0)

\thicklines
\dottedline{0.2}(0,5)(10,5)
\thinlines

\dashline{0.1}(0,6)(1,6)(1,1)
\dashline{0.1}(1,6)(2,6)(2,2)
\dashline{0.1}(2,5)(3,5)(3,2)
\dashline{0.1}(4,4)(4,5)(5,5)

\end{picture}
\end{center}
\caption{How to calculate the $n_{i}$'s}
\label{Calculation of n_i}
\end{figure}

The individual positions on the horizontal axis have not been displayed.
Instead, only the relevant intervals of them have been. For any $X\in (%
\mathbb{Z}
/p^{s}%
\mathbb{Z}
)^{r}$\ and any $a\in \{0,\ldots ,r\}$, the vertical bars in plain or dashed
lines and the horizontal dotted line above the $a$-th position show lower
bounds for the $p$-valuation of the $a$-th coefficient of $f(X)$. Thus as a
property of $D$, if $a\in K_{l}$, we have $v_{p}(f(X)_{a})\geq l$ as shown
by the plain line bars. Since the order of $f(X)$ is prescribed to be $%
p^{s-i}$, $v_{p}(f(X)_{a})\geq i$ as shown by the dotted line. Finally, we
put%
\begin{equation}
j=\min (k\in \{0,\ldots ,s\};\exists l\in K_{k},v_{p}(f(X)_{l})=i).
\end{equation}%
There exists such a $j$ ($j=2$ in the example on the graph) and if $a\leq
j-1 $, $v_{p}(f(X)_{a})>i$ as shown by the dashed line bars. These various
lower bounds partition $\{0,\ldots ,r\}$ into four subintervals
corresponding to the four factors in the above expression of $n_{i}$. The
sum amounts for all the possibilties for $j$.

For $i=0$, we have%
\begin{equation}
n_{0}=(p^{sr_{0}}-p^{(s-1)r_{0}})p^{s(r-r_{0})}=p^{sr}\left( 1-\frac{1}{%
p^{r_{0}}}\right) .
\end{equation}%
This quantity, which is a property of $M$, would increase strictly with $%
r_{0}$. Hence $r_{0}$ does not depend on the choice of $b$. For $i\geq 1$,
we suppose that for $j\leq i-1$, the $r_{j}$'s do not depend on $b$. Then
there exist a nonnegative integer $\alpha $ and a positive integer $\beta $
such that%
\begin{equation}
n_{i}=(\alpha p^{sr_{i}}+\beta (p^{sr_{i}}-p^{(s-1)r_{i}}))p^{s(r-\sigma
_{i-1}-r_{i})}=p^{s(r-\sigma _{i-1})}\left( \alpha +\beta -\frac{\beta }{%
p^{r_{i}}}\right) .
\end{equation}%
Again, we conclude that $r_{i}$ does not depend on $b$. The number $%
n_{i}^{\prime }$ of vectors of order $p^{s-i}$ in $M$, $i\in \{0,\ldots ,s\}$%
, would have been much more obvious a property of $M.$ But for $i\in
\{0,\ldots ,s-1\}$:%
\begin{multline}
n_{i}^{\prime }=\sum_{j=0}^{i}\left\{ \left[
\prod_{k=0}^{j-1}p^{((s-k-1)-(i-k))r_{k}}\right] \times \left[
p^{((s-j-1)-(i-j-1))r_{j}}-p^{((s-j-1)-(i-j))r_{j}}\right] \times \right. \\
\left. \times \left[ \prod_{k=j+1}^{i}p^{((s-k-1)-(i-k-1))r_{k}}\right]
\right\} \times \prod_{k=i+1}^{s-1}p^{((s-k-1)-(-1))r_{k}} \\
=\sum_{j=0}^{i}\left\{ \left[ \prod_{k=0}^{j-1}p^{(s-i-1)r_{k}}\right]
\times \left[ p^{(s-i)r_{j}}-p^{(s-i-1)r_{j}}\right] \times \left[
\prod_{k=j+1}^{i}p^{(s-i)r_{k}}\right] \right\} \times
\prod_{k=i+1}^{s-1}p^{(s-k)r_{k}},
\end{multline}%
with a cumbersome $\sum_{k=i+1}^{s-1}kr_{k}$\ appearing as an exponent in
the last factor. Even if we looked at $n_{i}^{\prime }/n_{i-1}^{\prime }$ to
handle the induction, that exponent would stay for the initialisation at $%
i=0 $.

Finally, harking back to the case where $d$ is not necessarily a power of a
prime, the number $r_{s}=r-(r_{0}+\ldots +r_{s-1})$ of diagonal entries of $%
\mathscr{D}(b)$ with $p$-valuation $v_{p}(d)$ is a property of $M$. We sum
up our results about simple reduction in the

\begin{theorem}
\label{Simple reduction theorem}For any rank-$r$\ submodule $M$ of $%
\mathbb{Z}
_{d}^{n}$, there exist a free basis $f$ of $%
\mathbb{Z}
_{d}^{n}$ and a minimal basis $b$ of $M$ such that:

\begin{enumerate}
\item $b$ is represented by a diagonal $n\times r$\ matrix $B$\ with respect
to $f$;

\item for all $i,j\in \{1,\ldots ,r\}$, $i<j$, we have $b_{ii}|b_{jj}$.
\end{enumerate}

\noindent Such a pair of bases $(f,b)$ can be found from any basis $b_{0}$
of $M$ by the simple reduction algorithm $\mathscr{D}$. Moreover, for any
pair $(f,b)$ as above, the sequence $(d/\nu (b_{ii}))_{i\in \{1,\ldots ,r\}}$
of the diagonal entries of $B$ "without unit factors" is the same and
therefore is a property of $M$. We shall call it the characteristic sequence
of $M$.
\end{theorem}

With the notations of the theorem, $M$ is free iff for all $i\in \{1,\ldots
,r\}$, $b_{ii}$ is a unit in $%
\mathbb{Z}
_{d}$, or in other words iff its characteristic sequence contains only $1$'s.

\begin{corollary}
\label{Incomplete family}Let $\beta _{0}=(b_{1},\ldots ,b_{r})$ be a free
family of $%
\mathbb{Z}
_{d}^{n}$. Then $r\leq n$ and there exist $n-r$ vectors $b_{r+1},\ldots
,b_{n}\in 
\mathbb{Z}
_{d}^{n}$ so that $\beta =(b_{1},\ldots ,b_{r},b_{r+1},\ldots ,b_{n})$ is a
free basis of $%
\mathbb{Z}
_{d}^{n}$.
\end{corollary}

\begin{proof}
Indeed, with $D$ the $(1,\ldots ,r;1,\ldots ,r)$ submatrix of the $r\times n$
diagonal matrix $\mathscr{D}(\beta _{0})$, a representative matrix for such
a $\beta $ with respect to the computational basis is%
\begin{equation}
L(\beta _{0})^{-1}\func{diag}(DR(\beta _{0})^{-1},I_{n-r}).
\end{equation}
\end{proof}

\begin{corollary}
\label{Characteristic sequence isomorphism}For any two submodules $M$ and $N$
of $%
\mathbb{Z}
_{d}^{n}$ with the same characteristic sequence, what implies that they have
the same rank, there exists an automorphism of $%
\mathbb{Z}
_{d}^{n}$\ that brings $M$ onto $N$.
\end{corollary}

\begin{proof}
Let $(f,b)$ (resp.~$(h,c)$) be a convenient pair for $M$ (resp.~$N$) as in
theorem~\ref{Simple reduction theorem}. Then the automorphism of $%
\mathbb{Z}
_{d}^{n}$\ defined by $b_{i}\mapsto c_{i}$, $i\in \{1,\ldots ,n\}$, brings $%
M $ onto $N$.
\end{proof}

\bigskip

The pair $(f,b)$ in theorem~\ref{Simple reduction theorem} is not unique.
For the sake of Section~\ref{A criterion for symplectic diagonalisation}, we
study the relation between the various suitable bases $f$'s. Let $%
(f^{(1)},b^{(1)})$ and $(f^{(2)},b^{(2)})$ be two convenient pairs and $P$
the $n\times n$ change of basis matrix defined by $f^{(1)}P=f^{(2)}$. Let us
work in a single Chinese factor. Let the $K_{i}$'s be defined as in (\ref{K
Definition}) plus%
\begin{equation}
K_{s}=\{r+1,\ldots ,n\}.  \label{Ks Definition}
\end{equation}%
For any $k\in \{0,\ldots ,n\}$, there exists some $i_{k}\in \{0,\ldots ,s\}$
so that $k\in K_{i_{k}}$. So $p^{i_{k}}f_{k}^{(2)}\in M$ and hence%
\begin{equation}
\forall i\in \{i_{k}+1,\ldots ,s\},\forall j\in K_{i},p^{i-i_{k}}|P_{jk}.
\label{Valuation in change of basis matrix}
\end{equation}%
Since $P$ is invertible, we also deduce from that latter result that for any 
$i\in \{0,\ldots ,s\}$, the $(K_{i};K_{i})$ diagonal block of $P$ is an
invertible matrix.

As a converse, for any convenient pair $(f,b)$\ and any invertible matrix $P$
satisfying relation (\ref{Valuation in change of basis matrix}), let $%
b^{\prime }$ be the family represented by the matrix $fP\mathscr{D}(b)$ and $%
N$ be the submodule of $M$ generated by $b^{\prime }$. Since $P$ is
invertible, $fP$ is a free family and $(fP,b^{\prime })$ is a convenient
pair for $N$. Hence $M$ and $N$ have the same characteristic sequence and
with the help of corollary~\ref{Characteristic sequence isomorphism}, we see
that they have the same cardinality. So $N=M$ and $(fP,b^{\prime })$ is a
convenient pair for $M$.

Let $\Sigma _{\mathscr{D}}(M)$ be the subgroup of $\func{GL}(n,%
\mathbb{Z}
_{d})$ that consists all the change of basis matrices we have just pointed
out.

\section{Symplectic reduction\label{Symplectic reduction Section}}

In this section, we replace $%
\mathbb{Z}
_{d}^{n}$ by $%
\mathbb{Z}
_{d}^{2n}$, that is to say we take an even number of copies of $%
\mathbb{Z}
_{d}$. Let $J=\left( 
\begin{array}{cc}
0 & 1 \\ 
-1 & 0%
\end{array}%
\right) $\ and $\omega $ be the canonical symplectic inner product in $%
\mathbb{Z}
_{d}^{2n}$. It is defined with respect to the canonical basis by the $%
2n\times 2n$\ block-diagonal\ matrix%
\begin{equation}
J_{n}=\left( 
\begin{array}{ccc}
J &  &  \\ 
& \ddots &  \\ 
&  & J%
\end{array}%
\right) .
\end{equation}%
A basis $(b_{1},\ldots ,b_{2n})$ such that for all $i,j\in \{1,\ldots ,n\}$, 
$i\neq j$,%
\begin{equation}
\omega (b_{2i-1},b_{2i})=-\omega (b_{2i},b_{2i-1})=1\text{ and\ }\omega
(b_{2i},b_{2j-1})=\omega (b_{2i},b_{2j})=0
\end{equation}%
is called a symplectic basis. The canonical basis is symplectic.

In simple reduction, we allowed any change of computational basis. In this
section, we are interested in reduction where changes of computational basis
can only be symplectic. This means that in the new basis, $\omega $ is still
to be represented by $J_{n}$.\ Matrices $L$\ used for left-mutiplication
thus have to satisfy the condition:%
\begin{equation}
L^{T}J_{n}L=J_{n},
\end{equation}%
where $L^{T}$ is the transpose of $L$. Such a matrix is called a symplectic
matrix. The identity matrix is symplectic. A matrix that represents a
symplectic basis with respect to another symplectic basis is symplectic.
Note that in $%
\mathbb{Z}
$, a symplectic matrix has determinant $\pm 1$. The same is thus true for a
symplectic matrix over $%
\mathbb{Z}
_{d}$. This proves that all symplectic matrices are invertible. Moreover,
the inverse of a symplectic matrix is symplectic. Our plan here is the same
as in the previous section. We first address reduction of a single vector
and afterwards that of a matrix. The case $n=1$ should be trivial to the
reader by now. Reduction of a single vector when $n\geq 2$\ relies itself on
the fondamental case $n=2$. The following substeps are elementary operations
that we shall use later on in the various steps of our symplectic reduction
algorithm for matrices. They form a sequence in order to reduce a vector
with four components $(x,y,z,t)^{T}$ using only symplectic changes of basis.

\bigskip

\textbf{Substep 1}: Let $x,y,z,t\in 
\mathbb{Z}
_{d}$ and $\delta =x\wedge y\wedge z\wedge t$. According to corollary~\ref%
{Invertible coeffs for gcd Cor.}, there exist $k_{1},k_{2},k_{3}\in 
\mathbb{Z}
_{d}$ and $u\in U(%
\mathbb{Z}
_{d})$\ such that%
\begin{equation}
\begin{array}{cccc}
\underbrace{\left( 
\begin{array}{cccc}
u & 0 & 0 & 0 \\ 
k_{1} & u^{-1} & k_{2} & k_{3} \\ 
-k_{3}u & 0 & 1 & 0 \\ 
k_{2}u & 0 & 0 & 1%
\end{array}%
\right) } & \left( 
\begin{array}{c}
x \\ 
y \\ 
z \\ 
t%
\end{array}%
\right) & = & \left( 
\begin{array}{c}
x_{1} \\ 
\delta \\ 
z_{1} \\ 
t_{1}%
\end{array}%
\right) \\ 
S_{1} &  &  & 
\end{array}%
\end{equation}%
where $x_{1},z_{1},t_{1}$ are byproducts of the choice of $k_{1},k_{2},k_{3}$
and$\ u$ and $S_{1}$ is symplectic.\quad $\blacklozenge $

\textbf{Substep 2}: Then, as in lemma~\ref{Reduction of a vector}, we find $%
v,w,k_{4},k_{5}\in 
\mathbb{Z}
_{d}$ such that%
\begin{equation}
\left\{ 
\begin{array}{l}
vz_{1}+wt_{1}=z_{1}\wedge t_{1}=z_{2} \\ 
-k_{5}z_{1}+k_{4}t_{1}=0 \\ 
vk_{4}+wk_{5}=1%
\end{array}%
\right. ,
\end{equation}%
and we perform a second left-multiplication:

\begin{equation}
\begin{array}{cccc}
\underbrace{\left( 
\begin{array}{cccc}
1 & 0 & 0 & 0 \\ 
0 & 1 & 0 & 0 \\ 
0 & 0 & v & w \\ 
0 & 0 & -k_{5} & k_{4}%
\end{array}%
\right) } & \left( 
\begin{array}{c}
x_{1} \\ 
\delta \\ 
z_{1} \\ 
t_{1}%
\end{array}%
\right) & = & \left( 
\begin{array}{c}
x_{1} \\ 
\delta \\ 
z_{2} \\ 
0%
\end{array}%
\right) , \\ 
S_{2} &  &  & 
\end{array}%
\end{equation}%
where $S_{2}$ is a symplectic matrix.\quad $\blacklozenge $

\textbf{Substep 3}: Since%
\begin{equation}
\delta =x\wedge y\wedge z\wedge t=x_{1}\wedge \delta \wedge z_{1}\wedge
t_{1}=x_{1}\wedge \delta \wedge z_{2},
\end{equation}%
we also have%
\begin{equation}
\delta \wedge z_{2}=(x_{1}\wedge \delta \wedge z_{2})\wedge z_{2}=\delta .
\end{equation}%
Thus we can find $k_{6}$ such that $k_{6}\delta +z_{2}=0$ and we perform a
third left-mutiplication:%
\begin{equation}
\begin{array}{cccc}
\underbrace{\left( 
\begin{array}{cccc}
1 & 0 & 0 & k_{6} \\ 
0 & 1 & 0 & 0 \\ 
0 & k_{6} & 1 & 0 \\ 
0 & 0 & 0 & 1%
\end{array}%
\right) } & \left( 
\begin{array}{c}
x_{1} \\ 
\delta \\ 
z_{2} \\ 
0%
\end{array}%
\right) & = & \left( 
\begin{array}{c}
x_{1} \\ 
\delta \\ 
0 \\ 
0%
\end{array}%
\right) , \\ 
S_{3} &  &  & 
\end{array}%
\end{equation}%
where $S_{3}$ is symplectic.\quad $\blacklozenge $

\bigskip

If $n>2$, we apply the process defined by this sequence of substeps $n-1$
times in order to end with a vector whose components are null except maybe
the first two ones. At step $i$, we set the $(2n+2-2i)$-th and the $%
(2n-2i+1) $-th components to $0$. For a single vector, we can go further and
set the second component to $0$ as in the second substep above. We shall
soon define a substep $4$ to complete this list of elementary operations.

\bigskip

It is in general not possible to diagonalise nor to trigonalise a matrix
using only a left-multiplication by a symplectic matrix. For instance, let
us try to do even weaker a job with the matrix $B$ in the following equality
over $%
\mathbb{Z}
/p^{s}%
\mathbb{Z}
$, $s\geq 1$:%
\begin{equation}
\begin{array}{cccc}
\underbrace{\left( 
\begin{array}{cccc}
\alpha & \ast & \gamma & k_{1} \\ 
\beta & \ast & \delta & k_{2} \\ 
0 & \ast & l_{1}p & \ast \\ 
0 & \ast & l_{2}p & \ast%
\end{array}%
\right) } & \underbrace{\left( 
\begin{array}{cc}
1 & 0 \\ 
0 & p \\ 
0 & 1 \\ 
0 & 0%
\end{array}%
\right) } & = & \left( 
\begin{array}{cc}
\alpha & \ast \\ 
\beta & \ast \\ 
0 & 0 \\ 
0 & 0%
\end{array}%
\right) . \\ 
L & B &  & 
\end{array}
\label{Impossibility example}
\end{equation}%
Our aim is to find a symplectic matrix $L$ so as to get rid of any nonzero
term in the last two rows. The first, third and fourth column vectors of $L$%
, let us call them $C_{1},C_{3}$ and $C_{4}$, must be as shown in (\ref%
{Impossibility example}). But as $L$ is supposed to be sympletic, $C_{3}$
must be free and $\omega (C_{1},C_{3})=0$. So there exist $k_{3},k_{4}\in 
\mathbb{Z}
_{d}$ such that $k_{3}\gamma +k_{4}\delta =1$ and $\alpha \delta =\beta
\gamma $. Hence $(\alpha ,\beta )$ is a multiple of $(\gamma ,\delta )$: 
\begin{subequations}
\begin{eqnarray}
\alpha &=&(k_{3}\gamma +k_{4}\delta )\alpha =(k_{3}\alpha +k_{4}\beta
)\gamma , \\
\beta &=&(k_{3}\gamma +k_{4}\delta )\beta =(k_{3}\alpha +k_{4}\beta )\delta .
\end{eqnarray}%
Since $C_{1}$ has to be free, $(k_{3}\alpha +k_{4}\beta )$ has to be a unit.
Then there exists $l\in 
\mathbb{Z}
_{d}$ such that 
\end{subequations}
\begin{equation}
\omega (C_{1},C_{4})=k_{2}\alpha -k_{1}\beta =(k_{3}\alpha +k_{4}\beta
)(k_{2}\gamma -k_{1}\delta )=(k_{3}\alpha +k_{4}\beta )(\omega
(C_{3},C_{4})-lp).
\end{equation}%
That quantity should be both $0$ and invertible and $L$ cannot be
symplectic. As for simple reduction, we shall make use of
right-multiplications to complete the reduction.\ Still, it is only possible
to lower-trigonalise that way. Despite that restrictive result, we are to
find another way of reducing that will prove sufficient to study Lagrangian
submodules in Section~\ref{Lagrangian submodules}. We shall also need the

\begin{criterion}
Let $a,x,y,z\in 
\mathbb{Z}
_{d}$ , $a\neq 0$, $x$ a multiple of $a$\ and%
\begin{equation}
m=\left( 
\begin{array}{cc}
a & x \\ 
0 & y \\ 
0 & z \\ 
0 & 0%
\end{array}%
\right) .
\end{equation}%
There exists a symplectic matrix $S$ such that $Sm$ is upper-triangular iff $%
z$ is multiple of $y$.
\end{criterion}

\begin{proof}
If $z$ is multiple of $y$, we can trigonalise $m$ by applying substep 3.

Given $a,x,y,z\in 
\mathbb{Z}
_{d}$ as specified in the criterion, $\delta =y\wedge z$ on the one hand and 
$k\in 
\mathbb{Z}
_{d},v\in U(%
\mathbb{Z}
_{d})$ on the other hand such that $\delta =ky+vz$, we have%
\begin{equation}
\begin{array}{cccc}
\underbrace{\left( 
\begin{array}{cccc}
1 & 0 & 0 & kv^{-1} \\ 
0 & 1 & 0 & 0 \\ 
0 & k & v & 0 \\ 
0 & 0 & 0 & v^{-1}%
\end{array}%
\right) } & \underbrace{\left( 
\begin{array}{cc}
a & x \\ 
0 & y \\ 
0 & z \\ 
0 & 0%
\end{array}%
\right) } & = & \underbrace{\left( 
\begin{array}{cc}
a & x \\ 
0 & y \\ 
0 & \delta \\ 
0 & 0%
\end{array}%
\right) }, \\ 
S_{4} & m &  & m^{\prime }%
\end{array}
\label{Substep 4}
\end{equation}%
where $S_{4}$ is symplectic. There exists $k^{\prime }\in 
\mathbb{Z}
_{d}$ such that $y=k^{\prime }\delta $ and let $\nu =\nu (\delta )$. In
order not to burden the argument with unessential details, we refer to the
Chinese remainder theorem to suppose that $d$ is a power of a prime, say $%
p^{s}$. Let $t=v_{p}(a)<s$. If $m^{\prime }$ is symplectically
trigonalisable as set out in the criterion, the symplectic matrix to use
must be as shown in the following equation:%
\begin{equation}
\begin{array}{cccc}
\underbrace{\left( 
\begin{array}{cccc}
w+k_{11}p^{s-t} & \ast & \ast & \ast \\ 
k_{21}p^{s-t} & w^{-1}+k_{22}p^{s-t} & k_{23}p^{s-t} & k_{24}p^{s-t} \\ 
k_{31}p^{s-t} & \alpha _{1} & -\alpha _{1}k^{\prime }+l_{1}\nu & \beta _{1}
\\ 
k_{41}p^{s-t} & \alpha _{2} & -\alpha _{2}k^{\prime }+l_{2}\nu & \beta _{2}%
\end{array}%
\right) } & \underbrace{\left( 
\begin{array}{cc}
a & x \\ 
0 & y \\ 
0 & \delta \\ 
0 & 0%
\end{array}%
\right) } & = & \left( 
\begin{array}{cc}
wa & \ast \\ 
0 & \ast \\ 
0 & 0 \\ 
0 & 0%
\end{array}%
\right) , \\ 
\text{\textit{symplectic}} & m^{\prime } &  & 
\end{array}%
\end{equation}%
with $w\in U(%
\mathbb{Z}
_{d})$. We leave the checking of that form to the reader. But the symplectic
inner product of the third and fourth columns of that matrix has to be $1$,
what proves with B\'ezout's theorem that $k^{\prime }$ and $\nu $ are
coprime. Let $\alpha ,\beta \in 
\mathbb{Z}
_{d}$ be such that $\alpha k^{\prime }+\beta \nu =1$. Then $\alpha y=\alpha
k^{\prime }\delta =(1-\beta \nu )\delta =\delta $.
\end{proof}

\bigskip

We can now state our

\bigskip

\textbf{Substep 4}: Let $x,y,z\in 
\mathbb{Z}
_{d}$, $\delta =y\wedge z$ and $X=(x,y,z,0)^{T}$ with respect to some
symplectic basis. One can find a new symplectic basis in which $X$ is
written $(x,y,\delta ,0)^{T}$. The way to do so is given in (\ref{Substep 4}%
).\quad $\blacklozenge $

\bigskip

In what follows, we shall need a refined version of the algorithm $%
\mathscr{A}$. Recall that for any $2n\times k$\ matrix $m$, $k\geq 1$, there
exists an $k\times k$ invertible matrix $R(m)$ such that $\mathscr{A}%
(m)=mR(m)$. For any $2n\times k$\ matrix $m$, $i\in \{1,\ldots ,2n\}$, $j\in
\{1,\ldots ,k-1\}$, and $m_{[i,j]}$ the $(i,\ldots ,2n;j,\ldots ,k)$\
submatrix of $m$, $\mathscr{A}_{i,j}$ will be the algorithm defined by%
\begin{equation}
\mathscr{A}_{i,j}(m)=m\left( 
\begin{array}{cc}
I_{j-1} & 0_{j-1,k-j+1} \\ 
0_{k-j+1,j-1} & R(m_{[i,j]})%
\end{array}%
\right) .
\end{equation}%
$\mathscr{A}_{i,j}$ does essentially the same job as $\mathscr{A}$ on
columns $j$ to $k$ of $m$, but it takes into account only the last $2n-i+1$
rows to maximise the order and combines those columns on the other lines
accordingly.

We now go on with the symplectic reduction algorithm for a single Chinese
factor. We suppose that $d=p^{s}$.

\bigskip

\textbf{Symplectic reduction algorithm} $\mathscr{S}$: Suppose we are given
a basis $b=(b_{1},\ldots ,b_{k})$\ of a submodule $M$ of $%
\mathbb{Z}
_{d}^{2n}$\ and $B$ is the matrix of size $2n\times k$\ whose $i$-th column
is $b_{i}$. To reduce $B$ in a symplectic way, the sarting point is $i=j=1$
and $B^{\prime }=B$, where $i$ and $j$ are some counters. Then while $i\leq
2n-3$\ and $j\leq k-1$, that is to say while there remain at least four
lines and two columns to deal with, do:

\begin{enumerate}
\item Apply $\mathscr{A}_{i,j}$ to $B^{\prime }$ and perform a first
left-multiplication by a symplectic matrix in order to set to $0$ all the
coefficients in the $j$-th column sarting from the $(i+1)$-th line. We
obtain a matrix $B^{(1)}$.

\item Apply $\mathscr{A}_{i+1,j+1}$\ to $B^{(1)}$ and perform a second
left-multiplication by a symplectic matrix to set to $0$ all the
coefficients in the $(j+1)$-th column sarting from the $(i+4)$-th line.
Indeed, as we see with the example above (equation \ref{Impossibility
example}), a step further as we planned to make it in the substeps could
affect the $j$-th column in a wrong way. We obtain a matrix $B^{(2)}$ whose $%
(i,\ldots ,i+3;j,j+1)$ submatrix is%
\begin{equation}
\left( 
\begin{array}{cc}
b_{i,j}^{(1)} & b_{i,j+1}^{(1)} \\ 
0 & b_{i+1,j+1}^{(1)} \\ 
0 & b_{i+2,j+1}^{(2)} \\ 
0 & b_{i+3,j+1}^{(2)}%
\end{array}%
\right) .
\end{equation}

\item Performing substeps 2 and 4, we get a matrix $B^{(3)}$ whose $%
(i,\ldots ,i+3;j,j+1)$ submatrix is of the form%
\begin{equation}
\left( 
\begin{array}{cc}
b_{i,j}^{(1)} & b_{i,j+1}^{(1)} \\ 
0 & xb_{i+2,j+1}^{(3)} \\ 
0 & b_{i+2,j+1}^{(3)} \\ 
0 & 0%
\end{array}%
\right) ,
\end{equation}%
with $x\in 
\mathbb{Z}
_{d}$. Notice the line index on the second line.

\item If $x$ is a unit, apply substep 3 to get a matrix $B^{(4)}$. If $x$ is
not a unit, just take $B^{(4)}=B^{(3)}$.

\item Since every coefficient in the $(i,\ldots ,2n;j+1,\ldots ,k)$
submatrix of $B^{(4)}$ is a multiple of $b_{i,j}^{(4)}=b_{i,j}^{(1)}$,
right-multiply $B^{(4)}$ by an appropriate, invertible matrix to set to $0$
the coefficients on the $i$-th row starting from the $(j+1)$-th column. We
obtain a matrix $B^{(5)}$.

\item If $x$ is a unit, since every coefficient in the $(i+1,\ldots
,2n;j+2,\ldots ,k)$ submatrix of $B^{(5)}$ is a multiple of $%
b_{i+1,j+1}^{(5)}=b_{i+1,j+1}^{(4)}$, right-multiply $B^{(5)}$\ by an
appropriate, invertible matrix to set to $0$ the coefficients on the $(i+1)$%
-th row starting from the $(j+2)$-th\ column. We obtain a new matrix $%
B^{\prime }$. If $x$ is not a unit, just take $B^{\prime }=B^{(5)}$.

\item If $x$ is a unit, increase $i$ and $j$ by $2$. If not, increase $i$ by 
$2$ and $j$ by $1$ only. In this latter case, we need not perform step 1 at
the next pass. The new $B^{(1)}$ is just the new $B^{\prime }$.
\end{enumerate}

\noindent Once this repeating process has ended, if $i=2n-1$, we reduce the
last two rows by means of a simple reduction, so as to have at most two
nonzero coefficients on them. If $i<2n-1\ $and $j=k$, apply a last
left-multilplication by a symplectic matrix to $B^{\prime }$\ so as to
reduce the last column as far as possible without modifying the others. As
an example, if we started with a $2n\times 2n$\ matrix $B$\ with $n=8$, we
may end up with a matrix of the form

\setstretch{0.7}%
\begin{equation}
\mathscr{S}(B)=\left( 
\begin{array}{cccccccccccccccc}
\ast &  &  &  &  &  &  &  &  &  &  &  &  &  &  &  \\ 
& \ast &  &  &  &  &  &  &  &  &  &  &  &  &  &  \\ \hline
&  & \ast &  &  &  &  &  &  &  &  &  &  &  &  &  \\ 
&  &  & \ast &  &  &  &  &  &  &  &  &  &  &  &  \\ \hline
&  &  &  & \ast &  &  &  &  &  &  &  &  &  &  &  \\ 
&  &  &  &  & R & \ast & \ast & \ast & \ast & \ast & \ast & \ast & \ast & 
\ast & \ast \\ \hline
&  &  &  &  & \ast &  &  &  &  &  &  &  &  &  &  \\ 
&  &  &  &  &  & \ast &  &  &  &  &  &  &  &  &  \\ \hline
&  &  &  &  &  &  & \ast &  &  &  &  &  &  &  &  \\ 
&  &  &  &  &  &  &  & R & \ast & \ast & \ast & \ast & \ast & \ast & \ast \\ 
\hline
&  &  &  &  &  &  &  & \ast &  &  &  &  &  &  &  \\ 
&  &  &  &  &  &  &  &  & \ast &  &  &  &  &  &  \\ \hline
&  &  &  &  &  &  &  &  &  & \ast &  &  &  &  &  \\ 
&  &  &  &  &  &  &  &  &  &  & \ast &  &  &  &  \\ \hline
&  &  &  &  &  &  &  &  &  &  &  & \ast &  &  &  \\ 
&  &  &  &  &  &  &  &  &  &  &  &  & \ast &  & 
\end{array}%
\right) ,  \label{Symplectic reduction Form}
\end{equation}%
\setstretch{1}%
where the meaning of the letter $R$ is explained below.$\quad \blacklozenge $

\bigskip

Horizontal lines of stars in (\ref{Symplectic reduction Form}) beginning
with an $R$ will be called rent lines and places marked with an $R$ rent
points. It is because of rent lines that we need actual
right-multiplications in steps 5 and 6 instead of merely setting some
coefficients to $0$ as in simple reduction. Without those
right-multiplications, we should not produce a basis matrix for the very
submodule we started from. A rent line can occur only on an even row.
Suppose $(i,j)$ is a rent point in the reduced matrix. Every coefficient in
the $(i,\ldots ,2n;j,\ldots ,k)$ submatrix is a multiple of the coefficient
underneath the rent point, at position $(i+1,j)$. So, if this coefficient is 
$0$, we may stop the algorithm. Last but not least about rents, it was
necessary to perform the algorithm in a single Chinese factor, since a rent
may occur at some position in some Chinese factor while not in another one.
This reduction procedure is thus linked in an essential way to the Chinese
remainder theorem.

The algorithm $\mathscr{S}$ consists in choosing basis vectors $f_{1},\ldots
,f_{2n}$ one after the other so as to obtain a basis matrix for $M$ of a
particular form with respect to the free basis $f$ thus constituted. But can
we avoid rents by a discerning choice of the $f_{i}$'s so as to get a
diagonal basis matrix for $M$? Is it a good strategy to choose a vector of
the greatest possible order as we did? If the issue of order has actually to
be addressed, is it of some use to discriminate between the vectors of a
given order? We shall answer those questions in Section~\ref{A criterion for
symplectic diagonalisation}, but we are now sufficiently provided to study
Lagrangian submodules.

\section{Lagrangian submodules\label{Lagrangian submodules}}

For any submodule $M$ of $%
\mathbb{Z}
_{d}^{2n}$, we define the symplectic orthogonal of $M$ by%
\begin{equation}
M^{\omega }=\{x\in 
\mathbb{Z}
_{d}^{2n};\forall y\in M,\omega (x,y)=0\}.
\end{equation}%
A submodule $M$ is called

\begin{itemize}
\item isotropic if $M\subset M^{\omega }$,

\item coisotropic if $M^{\omega }\subset M$,

\item symplectic if $M\cap M^{\omega }=\{0\}$,

\item Lagrangian if $M=M^{\omega }$.
\end{itemize}

\noindent Let $M$ be a Lagrangian submodule. $M$ is isotropic. Let us
suppose that there exists an isotropic submodule $N$ such that $M\subsetneq N
$. Then $M\subsetneq N\subset N^{\omega }\subset M^{\omega }$ and hence $M$
is not Lagrangian. Thus, a Lagrangian submodule is isotropic and maximal for
inclusion restricted to isotropic submodules. Theorem~\ref{Lagrangian
submodules Reduction} below will show that the converse is also true.

We are going to use symplectic reduction to find a very simple form for a
minimal basis matrix of $M$. As we saw it, we are to suppose that $d=p^{s}$.
Let $B_{0}$ be a basis matrix for $M$. The symplectic reduction $B=%
\mathscr{S}(B_{0})$ is still a basis matrix for $M$. Suppose some
coefficient appears on an even row, say at position $(2i,j)$, without a
rent. Since $M$ is isotropic, the symplectic product of the $(2i-1)$-th and
the $(2i)$-th column vectors of $\mathscr{S}(B)$ must be zero, what can be
written%
\begin{equation}
v_{p}(\mathscr{S}(B)_{2i-1,j-1})+v_{p}(\mathscr{S}(B)_{2i,j})\geq s.
\end{equation}%
The maximality of $M$ implies that this is in fact an equality. On the
contrary, if there is a rent point at position $(2i,j)$ and if the
coefficient of $\mathscr{S}(B)$\ at position $(2i-1,j-1)$\ has $p$-valuation 
$t$, then, by maximality of $M$, the vector%
\begin{equation}
C=(0,\ldots ,0,p^{s-t},0,\ldots ,0)^{T}
\end{equation}%
with $p^{s-t}$\ at the $(2i)$-th position, is in $M$. We insert this column
at position $2i$, that is to say between the $(2i-1)$-th and the $(2i)$-th
columns of $\mathscr{S}(B)$. Since $M$ is isotropic, every coefficient on
the $(2i)$-th line is a multiple of $p^{s-t}$ and we may set to $0$ every
coefficient on this line at right of the new column. We apply this trick to
each rent and obtain a diagonal matrix. So there exist $k\in \{1,\ldots ,n\}$
and $s_{1},\ldots ,s_{k}\in \{0,\ldots ,s\}$ so that the diagonal matrix%
\begin{equation}
D=\func{diag}(p^{s_{1}},p^{s-s_{1}},p^{s_{2}},p^{s-s_{2}},\ldots
,p^{s_{k}},p^{s-s_{k}})
\end{equation}%
is a basis matrix for $M$. If $k<n$, then $M$ would not be maximal. One
could add for instance the vector%
\begin{equation}
(0,\ldots ,0,1,0,\ldots ,0)^{T}
\end{equation}%
with $1$\ at the $(2k+1)$-th position and get a greater isotropic submodule.
So $k=n$. By construction of $\mathscr{S}(B)$, $s_{i}\leq s_{j}$ whenever $%
i<j$. Also note that $C$, as a vector of $M$, has to be a linear combination
of the column vectors of $\mathscr{S}(B)$. Since our trick to make good a
rent always yields a new basis matrix for $M$, the same is true for every
additional column. So, whether a diagonal coefficient of $D$ on an even row
appeared while dealing with a rent or not, our using of the algorithm $%
\mathscr{A}$ warrants that for all $i\in \{1,\ldots ,n\}$, $s_{i}\leq s-s_{i}
$. Since these results do not depend on the Chinese factor we chose, we have
proved the

\begin{theorem}
\label{Lagrangian submodules Reduction}Let $M$ be a submodule of $%
\mathbb{Z}
_{d}^{2n}$ and $d=\prod_{i\in I}p_{i}^{s_{i}}$ be the prime factor
decomposition of $d$. Then $M$ is Lagrangian iff the following two
conditions are satisfied. There exists a unique family%
\begin{equation}
(d_{1},\ldots ,d_{n})\in \left\{ 1,\ldots ,\prod\nolimits_{i\in
I}p_{i}^{\left\lfloor s_{i}/2\right\rfloor }\right\} ^{n}
\end{equation}%
such that $d_{1}|d_{2}|\ldots |d_{n}|d$ and there exists a $2n\times 2n$\
symplectic matrix $S$ such that%
\begin{equation}
S\times \func{diag}(d_{1},d/d_{1},d_{2},d/d_{2},\ldots ,d_{n},d/d_{n})
\end{equation}%
be a basis matrix for $M$.
\end{theorem}

As a remark to close this section, suppose the $(2i)$-th diagonal
coefficient of $D$, $i\in \{1,\ldots ,n-1\}$, appeared while applying the
algorithm $\mathscr{S}$ to $B_{0}$, that is to say there was no rent on the $%
(2i)$-th line. Then $s/2\geq s_{i+1}\geq s-s_{i}\geq s/2$ and so, for $j\geq
i$, $s_{j}=s/2$. If $s$ is odd, there is necessarily a rent on every even
row of $\mathscr{S}(B_{0})$ except the last one.

\section{A criterion for symplectic diagonalisation\label{A criterion for
symplectic diagonalisation}}

Lagrangian submodules are quite a particular case. In this section, we first
prove with an example that it is not always possible, for some submodule $M$%
, to find a symplectic basis $f$ and a $2n\times 2n$ diagonal matrix $D$
such that $fD$ be a basis of $M$. The diagonal entries of the $D$ need not
be arranged by increasing valuations. If such a pair $(f,D)$ exists, we
shall say that $M$ is nearly symplectic. Our aim will then be to provide an
criterion to know if a given $M$ is nearly symplectic. That will be done
with the algorithm $\mathscr{D}_{\omega }$ that also yields the symplectic
basis $f$ if any. We shall eventually see that as Lagrangian submodules,
symplectic ones form a particular kind of nearly symplectic submodules. For
the sake of simplicity, we take in this section $d=p^{s}$.

Let $c\in \{1,\ldots ,2n\}$ and $x=(x_{1},\ldots ,x_{c})$ a family of
vectors in $%
\mathbb{Z}
_{d}^{2n}$. The Gram matrix of $x$, $G=\func{Gram}(x)$, is the $c\times c$
matrix given by%
\begin{equation}
\forall i,j\in \{1,\ldots ,c\},g_{ij}=\omega (x_{i},x_{j}).
\end{equation}%
With matrices, if $B$ is the representative matrix of $x$ with respect to
the computational basis $e$, then $G=B^{T}J_{n}B$ and thus $G$ is
antisymmetric, but not necessarily invertible, even if $x$ is free. Yet, if $%
c=2n$ and $x$ is a free basis of $%
\mathbb{Z}
_{d}^{2n}$, then $B,G\in \func{GL}(2n,%
\mathbb{Z}
_{d})$. The discriminant of $x$ is the determinant of its Gram matrix:%
\begin{equation}
\Delta (x)=\det (\func{Gram}(x)).
\end{equation}%
Let $M$ be a submodule of $%
\mathbb{Z}
_{d}^{2n}$ and $F_{M}$ the set of all free bases $f$ of $%
\mathbb{Z}
_{d}^{2n}$ such that $M$ has a diagonal basis matrix with respect to $f$ as
in theorem~\ref{Simple reduction theorem}. We take the $K_{i}$'s, $i\in
\{0,\ldots ,s\}$, to be defined as in the proof and in the commentary of
that theorem in (\ref{K Definition}) and (\ref{Ks Definition})\footnote{%
Be careful that $n$ has been replaced by $2n$.}. Some of those intervals may
be empty. By restriction, the $K_{i}$'s determine a partition $K^{\prime }$
of $\{1,\ldots ,c\}$:%
\begin{equation}
\forall i\in \{0,\ldots ,s\},K_{i}^{\prime }=K_{i}\cap \{1,\ldots ,c\}.
\end{equation}%
For every $(i,j)\in \{0,\ldots ,s\}^{2}$, $G_{ij}$ will be the $%
(K_{i}^{\prime };K_{j}^{\prime })$ block of $G$. We also put $\widehat{G}%
_{ij}$ to be a matrix so that if $G_{ij}$ is not the empty matrix and if $%
s_{ij}=v_{p}(G_{ij})$, then $v_{p}\big(\widehat{G}_{ij}\big)=0$ and $%
p^{s_{ij}}\widehat{G}_{ij}=G_{ij}$. The matrix $\widehat{G}_{ij}$ thus
pointed out is not unique if $s_{ij}>0$. If $G_{ij}$ is the empty matrix,
then so is $\widehat{G}_{ij}$.

For now we take $c=2n$ only. A simplified study upon Gram matrices enables
us to give the simplest example of a non-nearly-symplectic submodule. The
pattern we catch a glimpse of here about those matrices will be seen in its
plain form afterwards. The reader who is interested only in the general case
may skip to the next part.

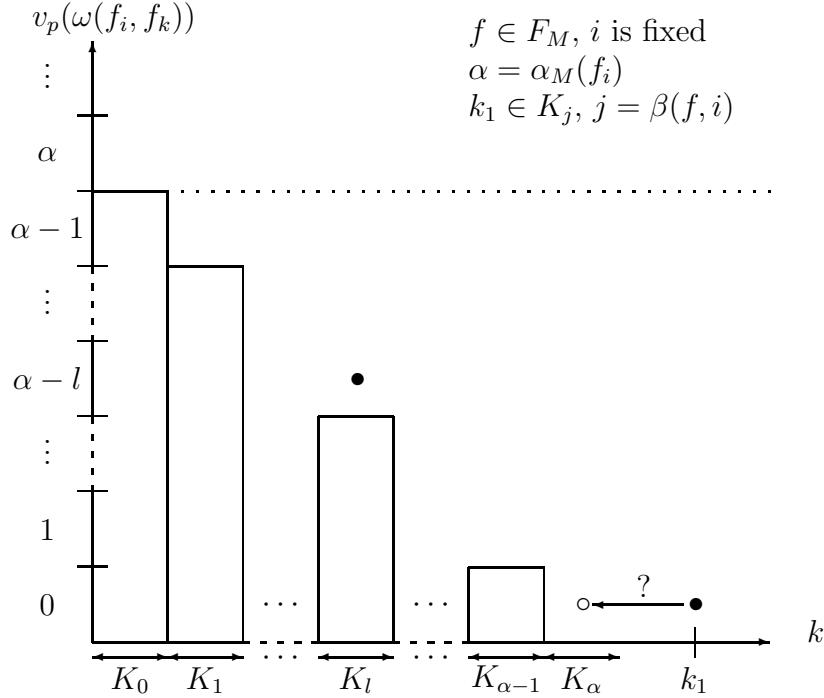
\begin{figure}[th]
\begin{center}
\begin{picture}(11,9.5)(-1,-0.5)

\put(0,0){\line(1,0){2}}
\dashline{0.1}(2,0)(3,0)
\put(2.25,0.4){$\cdots$}
\put(2.25,-0.3){$\cdots$}
\put(3,0){\line(1,0){1}}
\dashline{0.1}(4,0)(5,0)
\put(4.25,0.4){$\cdots$}
\put(4.25,-0.3){$\cdots$}
\put(5,0){\vector(1,0){4}}
\put(9.5,0){$k$}
\multiput(0.5,-0.2)(1,0){2}{\vector(1,0){0.5}}
\multiput(0.5,-0.2)(1,0){2}{\vector(-1,0){0.5}}
\put(3.5,-0.2){\vector(1,0){0.5}}
\put(3.5,-0.2){\vector(-1,0){0.5}}
\multiput(5.5,-0.2)(1,0){2}{\vector(1,0){0.5}}
\multiput(5.5,-0.2)(1,0){2}{\vector(-1,0){0.5}}
\multiputlist(0.5,-0.5)(1,0){$K_{0}$,$K_{1}$}
\multiputlist(3.5,-0.5)(1,0){$K_{l}$}
\multiputlist(5.5,-0.5)(1,0){$K_{\alpha -1}$,$K_{\alpha}$}
\put(8,-0.2){\rule{0.5pt}{0.4cm}}
\multiputlist(8,-0.5)(1,0){$k_{1}$}

\put(0,0){\line(0,1){2}}
\dashline{0.1}(0,2)(0,3)
\put(0,3){\line(0,1){1}}
\dashline{0.1}(0,4)(0,5)
\put(0,5){\vector(0,1){3}}
\multiput(-0.2,1)(0,1){7}{\rule{0.4cm}{0.5pt}}
\put(-0.8,8.2){$v_{p}(\omega (f_{i},f_{k}))$}
\multiputlist(-0.6,0.5)(0,1){$0$,$1$,$\raisebox{1.5ex}{\vdots}$,$\alpha -l$,$\raisebox{1.5ex}{\vdots}$,$\alpha -1$,$\alpha$,$\raisebox{1.5ex}{\vdots}$}

\drawline(0,6)(1,6)(1,0)
\drawline(1,5)(2,5)(2,0)
\drawline(3,0)(3,3)(4,3)(4,0)
\drawline(5,0)(5,1)(6,1)(6,0)

\thicklines
\dottedline{0.2}(0,6)(9,6)
\thinlines

\put(5,8){\makebox{$f\in F_{M}$, $i$ is fixed}}
\put(5,7.5){\makebox{$\alpha=\alpha_{M}(f_{i})$}}
\put(5,7){\makebox{$k_{1}\in K_{j},$ $j=\beta(f,i)$}}
\put(3.42,3.4){$\bullet$}
\put(7.92,0.4){$\bullet$}
\put(6.42,0.4){$\circ$}
\put(7.85,0.5){\vector(-1,0){1.2}}
\put(7.23,0.6){?}

\end{picture}
\end{center}
\caption{The functions $\protect\alpha _{M}$ and $\protect\beta $}
\label{Alpha and Beta}
\end{figure}

Let $f\in F_{M}$. For all $i\in \{1,\ldots ,2n\}$, we define%
\begin{subequations}%
\begin{align}
\alpha _{M}(f_{i})& =\min (v_{p}(\omega (f_{i},x));x\in M), \\
\beta (f,i)& =\min (j\in \{0,\ldots ,s\};\exists k\in K_{j},g_{ik}\in U(%
\mathbb{Z}
_{d})).
\end{align}%
\end{subequations}%
The graph on figure \ref{Alpha and Beta} illustrates the meaning of $\alpha
_{M}(f_{i})$ and $\beta (f,i)$. For any $k$ and $v$, a plain bullet at
position $(k,v)$ indicates that $v_{p}(g_{ik})=v$.

So there must exist $l\in \{0,\ldots ,\alpha \}$ and $k_{0}\in K_{l}$ so
that $v_{p}(g_{ik_{0}})=\alpha -l$. Let $i\in \{1,\ldots ,2n\}$, $j=\beta
(f,i)$ and $k_{1}\in K_{j}$ so that $g_{ik_{1}}\in U(%
\mathbb{Z}
_{d})$. Then $\alpha _{M}(f_{i})\leq v_{p}(\omega
(f_{i},p^{j}f_{k_{1}}))=j=\beta (f,i)$. This inequality is illustrated by
the second plain bullet at position $(k_{1},0)$.

We then consider a nearly symplectic submodule $M$ with a convenient pair $%
(f,D)$. If $(v_{p}(d_{ii}))_{i=1,\ldots ,2n}$\ is not an increasing
sequence, we use a $2n\times 2n$ permutation matrix $Q$\ so that the
diagonal coefficients of $Q^{T}DQ$ are arranged by increasing valuation. Let 
$f^{\prime }=fQ\in F_{M}$. On each line of $\func{Gram}(f^{\prime
})=Q^{T}J_{n}Q$, there is only one nonzero coefficient which is necessarily
invertible, in fact $1$ or $-1$, and it is clear that for all $i\in
\{1,\ldots ,2n\}$, $\alpha _{M}(f_{i}^{\prime })=\beta (f^{\prime },i)$. On
figure \ref{Alpha and Beta}, the equality $\alpha _{M}(f_{i})=\beta (f,i)$
is checked iff $k_{1}\in K_{\alpha }$.

We are now ready to find the annouced non-nearly-symplectic submodule. Let $%
s>1$ and $M$ be the submodule generated by the column vectors of the matrix $%
B$ in the following equation, with respect to $e$: 
\begin{equation}
\begin{array}{cccc}
\underbrace{\left( 
\begin{array}{cccc}
1 & 0 & 0 & 0 \\ 
0 & 1 & 1-p & 0 \\ 
0 & -1 & p & 0 \\ 
0 & 0 & 0 & 1%
\end{array}%
\right) } & \underbrace{\left( 
\begin{array}{cccc}
1 & 0 & 0 & 0 \\ 
0 & p & 0 & 0 \\ 
0 & 1 & 0 & 0 \\ 
0 & 0 & 0 & 0%
\end{array}%
\right) } & = & \left( 
\begin{array}{cccc}
1 & 0 & 0 & 0 \\ 
0 & 1 & 0 & 0 \\ 
0 & 0 & 0 & 0 \\ 
0 & 0 & 0 & 0%
\end{array}%
\right) \\ 
L & B &  & 
\end{array}%
\end{equation}%
We left-multiply $B$ by an invertible matrix $L$ so as to obtain a diagonal
matrix. Here, the diagonal coefficients of that latter matrix are already
arranged by increasing valuation. $K_{0}=\{1,2\}$ and $K_{s}=\{2,4\}$ are
the only nonempty intervals $K_{i}$. The new computational basis is $%
f=eL^{-1}\in F_{M}$ and the Gram matrix of $f$ is%
\begin{equation}
G=L^{-T}J_{n}L^{-1}=\left( 
\begin{array}{cc|cc}
0 & p & -1+p & 0 \\ 
-p & 0 & 0 & 1 \\ \hline
1-p & 0 & 0 & 1 \\ 
0 & -1 & -1 & 0%
\end{array}%
\right) ,
\end{equation}%
with $L^{-T}=(L^{-1})^{T}$. At the end of Section~\ref{Simple reduction
section}, we defined $\Sigma _{\mathscr{D}}(M)$. Here, any matrix $P\in
\Sigma _{\mathscr{D}}(M)$ is of the form%
\begin{equation}
P=\left( 
\begin{array}{cc}
A_{1} & A_{2} \\ 
0_{2,2} & A_{3}%
\end{array}%
\right) ,
\end{equation}%
with $A_{1},A_{3}\in \func{GL}(2,%
\mathbb{Z}
_{d})$. The Gram matrix of $f^{\prime }=fP$ is of the form%
\begin{equation}
P^{T}GP=\left( 
\begin{array}{cc}
pA_{1}^{T}\widehat{G}_{00}A_{1} & A_{4} \\ 
-A_{4} & A_{5}%
\end{array}%
\right) ,\text{ with }\widehat{G}_{00}=\left( 
\begin{array}{cc}
0 & 1 \\ 
-1 & 0%
\end{array}%
\right)
\end{equation}%
and $A_{1}^{T}\widehat{G}_{00}A_{1},A_{4}\in \func{GL}(2,%
\mathbb{Z}
_{d})$. But we see that for any $i\in K_{0}=\{1,2\}$, $\alpha
_{M}(f_{i}^{\prime })=1<\beta (f^{\prime },i)=s$. Comparing to the result of
the previous paragraph, this proves our claim that $M$ is not nearly
symplectic.

What if $s=1$? In that case, the matrix obtained by swapping the second and
third columns of $B$, namely $\func{diag}(1,0,1,0)$, is a convenient
diagonal basis matrix for $M$ with respect to the symplectic basis $e$.

\bigskip

In the remaining part of this section, we tell how to know whether a given
submodule $M$ is nearly symplectic or not and how to find a convenient pair $%
(f,D)$. We shall need a little more vocabulary. Let $b$ be a free basis of $%
\mathbb{Z}
_{d}^{2n}$, $\sigma \in \mathfrak{S}_{2n}$ a permutation of $\{1,\ldots
,2n\} $ and $Q$ the representative matrix of $\sigma $, that is to say the
only nonzero coefficients of $Q$ are equal to $1$ and are located at the
positions $(i,\sigma (i))_{i=1,\ldots ,2n}$. We denote $b_{\sigma }$ the
free basis $(b_{\sigma (1)},\ldots ,b_{\sigma (2n)})$ of $%
\mathbb{Z}
_{d}^{2n}$ and say that $b$ is $\sigma $-symplectic if $b_{\sigma }=bQ^{T}$
is symplectic. In that case, the representative matrix of $\omega $ in basis 
$b$ is $Q^{T}J_{n}Q$. A \ $2n\times 2n$\ matrix $L$ is said $\sigma $%
-symplectic if $QLQ^{T}$ is symplectic or equivalently if%
\begin{equation}
L^{T}(Q^{T}J_{n}Q)L=Q^{T}J_{n}Q.
\end{equation}%
Thus the conjugation by $L$ preserves the matrix representative of $\omega $
in $b$, $L$ is invertible and $L^{-1}$ is still $\sigma $-symplectic. If $b\ 
$and $L$ are $\sigma $-symplectic, $bL=bQ^{T}QL$ is still a $\sigma $%
-symplectic basis and if $B$ is the representative matrix of $b$ with
respect to a $\sigma $-symplectic basis $f$, then $B$ is a $\sigma $%
-symplectic matrix. Indeed, $fQ^{T}\ $and $bQ^{T}=(fQ^{T})(QBQ^{T})$ are
symplectic bases and hence $QBQ^{T}$ is a symplectic matrix.

The notions of scalar and set fringe we are going to define involve the $%
K_{i}$'s and thus are meanigless unless a reference submodule or a suitable
partition of $\{1,\ldots ,2n\}$\ is specified. Let $M$ be a submodule of $%
\mathbb{Z}
_{d}^{2n}$. Define the $K_{i}$'s accordingly and let $\kappa $ be the map%
\begin{equation}
\begin{array}{crcll}
\kappa : & \{1,\ldots ,2n\} & \longrightarrow & \{0,\ldots ,s\}, & \text{%
such that }i\in K_{\kappa (i)}. \\ 
& i & \longmapsto & \kappa (i) & 
\end{array}%
\end{equation}%
Then for any Gram matrix $G$ of size $\leq 2n$ and containing at least one
unit, we define the scalar ($M$-)fringe of $G$ by%
\begin{equation}
\func{fr}_{M}(G)=\min (\kappa (i)+\kappa (j);g_{ij}\in U(%
\mathbb{Z}
_{d}))
\end{equation}%
or equivalently%
\begin{equation}
\func{fr}_{M}(G)=\min (i+j;v_{p}(G_{ij})=0).
\end{equation}%
The ($M$-)fringe of $G$ is the set of all coefficients $g_{ij}$ such that $%
\kappa (i)+\kappa (j)\leq \func{fr}_{M}(G)$. A block $G_{ij}$ is said to be
in the fringe of $G$ if $\gamma _{ij}=\func{fr}_{M}(G)-i-j\geq 0$. Whenever
all the blocks $G_{ij}$ in the fringe of $G$ verify $v_{p}(G_{ij})\geq
\gamma _{ij}$, we shall say that the ($M$-)fringe of $G$\ is good. If there
exists $(i,j)\in \{1,\ldots ,2n\}^{2}$ such that 
\begin{equation}
g_{ij}\in U(%
\mathbb{Z}
_{d})\text{ with }\gamma _{\kappa (i)\kappa (j)}=0,
\label{Unit on the fringe}
\end{equation}%
and 
\begin{subequations}
\label{Nice fringe conditions}
\begin{align}
\forall k\leq i,& \ v_{p}(g_{kj})\geq \gamma _{\kappa (k)\kappa (j)}, \\
\forall l\leq j,& \ v_{p}(g_{il})\geq \gamma _{\kappa (i)\kappa (l)},
\end{align}%
we shall say that the ($M$-)fringe of $G$\ is nice. Of course a good $M$%
-fringe is a nice $M$-fringe. Let us give an example. If a block $G_{ij}$
with $i+j=3$ contains a unit, then the following Gram matrix has a good
fringe and scalar fringe $3$. 
\end{subequations}
\begin{equation}
G=\left( 
\begin{array}{ccccc}
\cline{1-4}
\multicolumn{1}{|c}{p^{3}\widehat{G}_{00}} & \multicolumn{1}{|c}{p^{2}%
\widehat{G}_{01}} & \multicolumn{1}{|c}{p\widehat{G}_{02}} & 
\multicolumn{1}{|c}{G_{03}} & \multicolumn{1}{|c}{\cdots \rule{0cm}{5.5mm}}
\\ \cline{1-4}
\multicolumn{1}{|c}{p^{2}\widehat{G}_{10}} & \multicolumn{1}{|c}{p\widehat{G}%
_{11}} & \multicolumn{1}{|c}{G_{12}} & \multicolumn{1}{|c}{} & \rule%
{0pt}{5.5mm} \\ \cline{1-3}
\multicolumn{1}{|c}{p\widehat{G}_{20}} & \multicolumn{1}{|c}{G_{21}} & 
\multicolumn{1}{|c}{} &  & \rule{0pt}{5.5mm} \\ \cline{1-2}
\multicolumn{1}{|c}{G_{30}} & \multicolumn{1}{|c}{} &  &  & \rule{0pt}{5.5mm}
\\ \cline{1-1}
\vdots &  &  &  & 
\end{array}%
\right) .  \label{Good fringe Example}
\end{equation}%
We shall need the following lemma and corollary.

\begin{lemma}
\label{Preserving scalar and good fringe}Let $M$ be a submodule in $%
\mathbb{Z}
_{d}^{2n}$. Let $b$ be a free basis of $%
\mathbb{Z}
_{d}^{2n}$ with Gram matrix $G$ and assume that $G$ has a good $M$-fringe.
Then for any $P\in \Sigma _{\mathscr{D}}(M)$, $P^{T}GP$ has a good $M$%
-fringe with the same scalar $M$-fringe as $G$.
\end{lemma}

That is to say the form (\ref{Good fringe Example}), with the particular
scalar $M$-fringe required, is preserved under conjugation by a matrix in $%
\Sigma _{\mathscr{D}}(M)$.

\begin{proof}
The reference submodule is $M$. Let $H=GP$. For every block $H_{ij}$ of $H$,
we have%
\begin{equation}
H_{ij}=\sum_{k=0}^{j-1}G_{ik}P_{kj}+G_{ij}P_{jj}+%
\sum_{k=j+1}^{s}G_{ik}P_{kj}.
\end{equation}%
As to the first sum, for every $k\in \{0,\ldots ,j-1\}$, we have%
\begin{equation}
v_{p}(G_{ik})+v_{p}(P_{kj})\geq v_{p}(G_{ik})\geq \gamma _{ik}\geq \gamma
_{ij}+1
\end{equation}%
and we refer to relations (\ref{Valuation and sum}) and (\ref{Valuation and
product}) of Appendix~\ref{CRT} to see that 
\begin{equation}
v_{p}(G_{ik}P_{kj})\geq \min (v_{p}(G_{ik})+v_{p}(P_{kj}),s)\geq \gamma
_{ij}+1.
\end{equation}%
Since $P_{jj}$ is invertible, the lines of $G_{ij}P_{jj}\ $are of the same
order as the lines of $G_{ij}$ respectively and then%
\begin{equation}
v_{p}(G_{ij}P_{jj})=v_{p}(G_{ij})\geq \gamma _{ij}.
\end{equation}%
As to the second sum, for every $k\in \{j+1,\ldots ,s\}$, the inequality%
\begin{equation}
v_{p}(G_{ik})+v_{p}(P_{kj})\geq (\func{fr}(G)-i-k)+(k-j)=\gamma _{ij}
\label{Preserving fringe Second sum}
\end{equation}%
implies that%
\begin{equation}
v_{p}(G_{ik}P_{kj})\geq \gamma _{ij}.
\end{equation}%
So $v_{p}(H_{ij})\geq \gamma _{ij}$. Let $(i,j)$ be such that $\gamma
_{ij}=0 $ and $v_{p}(G_{ij})=0$. Then the inequality in (\ref{Preserving
fringe Second sum}) may be modified as%
\begin{equation}
\forall k\in \{j+1,\ldots ,s\},v_{p}(G_{ik})+v_{p}(P_{kj})\geq 0+(k-j)\geq 1,
\end{equation}%
and we see that $v_{p}(H_{ij})=0$. So $H$ has a good fringe with scalar
fringe $\func{fr}(G)$. In the same manner, $P^{T}GP=P^{T}H$ has a good
fringe with scalar fringe $\func{fr}(G)$.
\end{proof}

\begin{corollary}
\label{Nearly symplecticity and fringe}Let $M$ be a nearly symplectic
submodule of $%
\mathbb{Z}
_{d}^{2n}$ and $f\in F_{M}$. Then the matrix $\func{Gram}(f)$ has a good $M$%
-fringe.
\end{corollary}

\begin{proof}
By assumption, there exists $\sigma \in \mathfrak{S}_{2n}$ and $f^{\prime }$
a $\sigma $-symplectic basis in $F_{M}$. We have already seen that $%
G^{\prime }=\func{Gram}(f^{\prime })$ has a good fringe (with respect to any
submodule). Besides, there exists $P\in \Sigma _{\mathscr{D}}(M)$ so that $%
f=f^{\prime }P$. So, $\func{Gram}(f)=P^{T}G^{\prime }P$ has a good $M$%
-fringe.
\end{proof}

\bigskip

We can now give the algorithm for symplectic diagonalisation whenever
possible:

\bigskip

\textbf{Algorithm} $\mathscr{D}_{\omega }$: Let $M$ be a submodule of $%
\mathbb{Z}
_{d}^{2n}$, $b$ a basis of $M$ and $B$ its representative basis matrix with
respect to any computational basis $e^{\prime }$.

Let $f=e^{\prime }L(B)^{-1}\in F_{M}$, where $L(B)$ was defined within the
algorithm $\mathscr{D}_{0}$ (see page~\pageref{L in algorithm D0}), $%
\widetilde{M}=M$ and $b^{\prime }$ be the empty sequence with values in $%
\mathbb{Z}
_{d}^{2n}$. Let also $c$ be a counter with initial value $0$.

While $G=\func{Gram}(f)$ has a nice $\widetilde{M}$-fringe, do

\begin{enumerate}
\item Choose a pair $(i,j)\in \{1,\ldots ,2n-2c\}^{2}$ that verifies
conditions (\ref{Unit on the fringe}) and (\ref{Nice fringe conditions}) and
perform the partial Gram-Schmidt orthogonalisation process: 
\begin{subequations}
\begin{gather}
f_{i}^{\prime }=f_{i},\quad f_{j}^{\prime }=f_{j}, \\
\forall k\in \{1,\ldots ,2n-2c\}\setminus \{i,j\},f_{k}^{\prime
}=f_{k}-g_{ij}^{-1}g_{\rule{0pt}{7.2pt}ik}f_{j}+g_{ij}^{-1}g_{\rule%
{0pt}{7.2pt}jk}f_{i}.
\end{gather}%
Owing to the nice fringe condition, the corresponding change of basis matrix 
$R$ is in $\Sigma _{\mathscr{D}}(M)$. With $i\leq j$ and $g_{ij}=1$, it
reads 
\end{subequations}
\begin{equation}
R=\left( 
\begin{array}{ccccccccccc}
1 &  &  &  &  &  &  &  &  &  &  \\ 
& \ddots &  &  &  &  &  &  &  &  &  \\ 
&  & 1 &  &  &  &  &  &  &  &  \\ \hline
g_{j1} & \cdots & g_{j,i-1} & 1 & g_{j,i+1} & \cdots & g_{j,j-1} & 0 & 
g_{j,j+1} & \cdots & g_{j,2n} \\ \hline
&  &  &  & 1 &  &  &  &  &  &  \\ 
&  &  &  &  & \ddots &  &  &  &  &  \\ 
&  &  &  &  &  & 1 &  &  &  &  \\ \hline
-g_{i1} & \cdots & -g_{i,i-1} & 0 & -g_{i,i+1} & \cdots & -g_{i,j-1} & 1 & 
-g_{i,j+1} & \cdots & -g_{i,2n} \\ \hline
&  &  &  &  &  &  &  & 1 &  &  \\ 
&  &  &  &  &  &  &  &  & \ddots &  \\ 
&  &  &  &  &  &  &  &  &  & 1%
\end{array}%
\right) ,
\end{equation}%
where the two special rows are the $i$-th one and the $j$-th one
respectively. For any $k\in \{1,\ldots ,2n-2c\}\setminus \{i,j\}$, $%
f_{k}^{\prime }\in \left\langle f_{i}^{\prime },f_{j}^{\prime }\right\rangle
^{\omega }$ and since $R\in \Sigma _{\mathscr{D}}(M)$, $f^{\prime }=fR\in
F_{M}$.

\item Let $b^{\prime }$ be the concatenation of $b^{\prime }$ and $%
(g_{ij}^{-1}f_{i}^{\prime },f_{j}^{\prime })$.

\item Rename $\widetilde{M}\cap \left\langle f_{i}^{\prime },f_{j}^{\prime
}\right\rangle ^{\omega }$ as $\widetilde{M}$.

\item Rename $f^{\prime }\setminus \{f_{i}^{\prime },f_{j}^{\prime }\}$ as $%
f $.

\item Increase $c$ by $1$.
\end{enumerate}

Whenever $\mathscr{D}_{\omega }(b)=\mathscr{D}_{\omega }(e^{\prime
},B)=b^{\prime }$ has cardinality $2n$, then it is a symplectic basis of $%
\mathbb{Z}
_{d}^{2n}$, $M$ is nearly symplectic and there exists $\sigma \in \mathfrak{S%
}_{2n}$\ so that $b_{\sigma }^{\prime }\in F_{M}$.\quad $\blacklozenge $

\bigskip

Since $e^{\prime }L(B)^{-1}$ is a free basis of $%
\mathbb{Z}
_{d}^{2n}$, its Gram matrix is invertible and thus has a well-defined $M$%
-fringe.\ Then the discriminant%
\begin{equation}
\Delta (f^{\prime }\setminus \{f_{i}^{\prime },f_{j}^{\prime }\})=\pm \Delta
(f^{\prime })=\pm \det (R)^{2}\Delta (f)
\end{equation}%
being a unit, all the forthcoming matrices $G$ have a well-defined $%
\widetilde{M}$-fringe and $\mathscr{D}_{\omega }$ is a valid algorithm. Now
if $M$ is nearly symplectic, does this algorithm yields the matrix $%
b^{\prime }$ we search for? Besides, in step 1, the pair $(i,j)$ is not
unique. So we are to prove that if $M$ is nearly symplectic, the algorithm $%
\mathscr{D}_{\omega }$, with any choice of the pairs, builds a symplectic
basis $b^{\prime }$ that endows $M$ with a diagonal basis matrix.

Let $M$ be a nearly symplectic submodule of $%
\mathbb{Z}
_{d}^{2n}$, $\sigma \in \mathfrak{S}_{2n}$ with representative matrix $Q$
and $h$ a $\sigma $-symplectic basis in $F_{M}$. Let also $B$ be a basis
matrix of $M$ with respect to any computational basis $e^{\prime }$ and let
us carry out the algorithm. Corollary~\ref{Nearly symplecticity and fringe}
shows that the first Gram matrix $G$ has a nice $M$-fringe. We choose a
convenient pair $(i,j)$ and find the first two vectors of $b^{\prime }$ by
performing steps 1 and 2. Then with $N=M\cap \left\langle f_{i}^{\prime
},f_{j}^{\prime }\right\rangle ^{\omega }$, we want to show that the Gram
matrix of $f^{\flat }=f^{\prime }\setminus \{f_{i}^{\prime },f_{j}^{\prime
}\}$ has a nice $N$-fringe. From now on, we consider $N$ as a submodule of $%
\left\langle f^{\flat }\right\rangle $ exclusively. With that convention, $%
f^{\flat }\in F_{N}$ and corollary~\ref{Nearly symplecticity and fringe}
tells us that it suffices to show that $N$ is nearly symplectic.

There exists $P\in \Sigma _{\mathscr{D}}(M)$ such that $f^{\prime }=hP$. For
any $m\in \{1,\ldots ,2n\}$, let $\ell (m)$ be the index defined by $\omega
(h_{m},h_{\ell (m)})=\pm 1$. Since $\func{fr}_{M}(h)=\func{fr}_{M}(f^{\prime
})$ as shown by lemma~\ref{Preserving scalar and good fringe}, we have%
\begin{equation}
\kappa (m)+\kappa (\ell (m))\geq \kappa (i)+\kappa (j)
\label{Sigma Fringe lower-bound}
\end{equation}%
and hence 
\begin{subequations}
\begin{align}
\kappa (m)<\kappa (i)& \Rightarrow \kappa (\ell (m))>\kappa (j), \\
\kappa (m)<\kappa (j)& \Rightarrow \kappa (\ell (m))>\kappa (i).
\end{align}%
So, and because $\omega (f_{i}^{\prime },f_{j}^{\prime })=g_{ij}$ is a unit,
there exist $k\in K_{\kappa (i)}$ and $l=\ell (k)\in K_{\kappa (j)}$ so that
the coefficients $p_{ki}$ and $p_{lj}$ in $P$ are units. So $QP$ has a unit
on its $\sigma ^{-1}(k)$-th line. Let $L$ be a symplectic matrix so that $%
LQP $ has all but its $\sigma ^{-1}(k)$-th coefficient equal to $0$. Since
we suppose we know where an invertible coefficient is in the $i$-th column
of $P $, the substeps of the symplectic reduction algorithm are unuseful to
find $L $. Instead, we form a symplectic matrix inspired by the Gaussian
reduction. For instance, if $\sigma ^{-1}(k)=1$, then $\sigma ^{-1}(l)=2$
and $L$ is of the form 
\end{subequations}
\begin{equation}
L=\left( 
\begin{array}{ccccccc}
1 &  &  &  &  &  &  \\ 
k_{0} & 1 & -k_{4} & k_{3} & \cdots & -k_{2n} & k_{2n-1} \\ 
k_{3} &  & 1 &  &  &  &  \\ 
k_{4} &  &  & 1 &  &  &  \\ 
\vdots &  &  &  & \ddots &  &  \\ 
k_{2n-1} &  &  &  &  & 1 &  \\ 
k_{2n} &  &  &  &  &  & 1%
\end{array}%
\right) .
\end{equation}%
Then the $i$-th column of $P^{\prime }=Q^{T}LQP$ has all but its $k$-th
coefficient equal to $0$. The basis $h^{\prime }=hQ^{T}L^{-1}Q$ is still $%
\sigma $-symplectic and $f^{\prime }=hP=h^{\prime }P^{\prime }$. Moreover,
the matrix $Z=Q^{T}LQ$ is in $\Sigma _{\mathscr{D}}(M)$ and thus $P^{\prime
}\in \Sigma _{\mathscr{D}}(M)$. Indeed, the coefficients in the $k$-th
column of $Z$ have the right valuations by construction. The coefficients on
the $l$-th row not in the $k$-th nor in the $l$-th columns were determined
so that $Z$ is $\sigma $-symplectic. In particular:%
\begin{equation}
\forall m\in \{1,\ldots ,l-1\}\setminus \{k\},\omega (Z_{m},Z_{k})=\pm
z_{\ell (m),k}\pm z_{lm}=0,
\end{equation}%
where for all $i$, $Z_{i}$ is the $i$-th column vector of $Z$. And according
to relation (\ref{Sigma Fringe lower-bound}),%
\begin{equation}
\forall m\in \{1,\ldots ,l-1\}\setminus \{k\},v_{p}(z_{\ell (m),k})\geq
\kappa (\ell (m))-\kappa (i)\geq \kappa (l)-\kappa (m).
\end{equation}%
Thus $v_{p}(z_{lm})\geq \kappa (l)-\kappa (m)$ and that proves that $Z\in
\Sigma _{\mathscr{D}}(M)$. The coefficient $p_{lj}^{\prime }$ divides $%
g_{ij} $ and hence is a unit. So we apply the same kind of reduction as
before to the $j$-th column of $P^{\prime }$ while preserving the $i$-th one
and find a $\sigma $-symplectic basis $h^{\prime \prime }$ and an invertible
matrix $P^{\prime \prime }\in \Sigma _{\mathscr{D}}(M)$ so that $f^{\prime
}=h^{\prime \prime }P^{\prime \prime }$. We may suppose without loss of
generality that $g_{ij}=1$. Then the vectors $h_{k}^{\prime \prime }$ and $%
h_{l}^{\prime \prime }$ may be redefined under a multiplication by a unit
factor so that $p_{ki}^{\prime \prime }=p_{lj}^{\prime \prime }=1$. If we
assume that $i<j$ and $k<l$ for instance, $P^{\prime \prime }$ is of the form%
\begin{equation}
\makebox[0pt][l]{$\hspace{1.05cm}
\begin{array}{c}
\left( \begin{array}{c}
\rule{0pt}{2.4cm} \rule{2.45cm}{0pt}
\end{array}\right) \\
\rule{0pt}{9.8mm}
\end{array}
$}%
\begin{array}{cccccccc}
& \ast & \multicolumn{1}{|c}{} & \multicolumn{1}{|c}{\ast} & 
\multicolumn{1}{|c}{} & \multicolumn{1}{|c}{\ast} &  &  \\ \cline{2-6}
&  & \multicolumn{1}{|c}{1} & \multicolumn{1}{|c}{} & \multicolumn{1}{|c}{0}
& \multicolumn{1}{|c}{} & \hspace*{0.4cm}\leftarrow & k \\ \cline{2-6}
P^{\prime \prime }=\hspace*{0.3cm} & \ast & \multicolumn{1}{|c}{} & 
\multicolumn{1}{|c}{\ast} & \multicolumn{1}{|c}{} & \multicolumn{1}{|c}{\ast}
&  &  \\ \cline{2-6}
&  & \multicolumn{1}{|c}{0} & \multicolumn{1}{|c}{} & \multicolumn{1}{|c}{1}
& \multicolumn{1}{|c}{} & \hspace*{0.4cm}\leftarrow & l \\ \cline{2-6}
& \ast & \multicolumn{1}{|c}{} & \multicolumn{1}{|c}{\ast} & 
\multicolumn{1}{|c}{} & \multicolumn{1}{|c}{\ast} &  &  \\ 
&  & \uparrow &  & \uparrow & \rule{0cm}{0.5cm} &  &  \\ 
&  & i &  & j &  &  & 
\end{array}%
\end{equation}%
Let $h^{\flat }=h^{\prime \prime }\setminus \{h_{i}^{\prime \prime
},h_{j}^{\prime \prime }\}$ and $P^{\flat }$ be the matrix obtained by
deleting the $k$-th and $l$-th rows as well as the $i$-th and $j$-th columns
of $P^{\prime \prime }$. Now $f_{i}^{\prime }=h_{k}^{\prime \prime }$ and $%
f_{j}^{\prime }=h_{l}^{\prime \prime }$ so that $f^{\flat }=h^{\flat
}P^{\flat }$. By construction, $P^{\flat }\in \Sigma _{\mathscr{D}}(N)$. So $%
h^{\flat }\in F_{N}$. But since $h^{\prime \prime }$ is $\sigma $-symplectic
and $\omega (h_{k}^{\prime \prime },h_{l}^{\prime \prime })=1$, there exists 
$\rho \in \mathfrak{S}_{2n-2}$ such that $h^{\flat }$ is $\rho $-symplectic.
That proves that $N$ is nearly symplectic.

\bigskip

We end this section with a proposition that shows the difference between
symplectic and nearly symplectic submodules.

\begin{proposition}
Let $M$ be a submodule of $%
\mathbb{Z}
_{d}^{2n}$. Then $M$ is symplectic iff $M$ is nearly symplectic and such
that $M+M^{\omega }=%
\mathbb{Z}
_{d}^{2n}$. In that case, $M$ is free and of even rank.
\end{proposition}

\begin{proof}
If $M=\{0\}$, both terms of the equivalence are checked and $M$ is obviously
free and of even rank. So let $M$ be a nonzero symplectic submodule and let $%
f\in F_{M}$. Since $p^{s-1}f_{1}\in M\setminus \{0\}$ and $M\cap M^{\omega
}=\{0\}$, there exists $x=\sum_{i=2}^{2n}x_{i}f_{i}\in M$ such that $\omega
(p^{s-1}f_{1},x)\neq 0$. Thus $x$ is free, $\omega (f_{1},x)$ is a unit,
there exists $j\in K_{0}\setminus \{1\}$ so that $\omega (f_{1},f_{j})$ is a
unit and $f_{1}\in M$. That proves that $\func{Gram}(f)$ has a good fringe.
We then perform the partial Gram-Schmidt process and find a new basis $%
f^{\prime }\in F_{M}$: 
\begin{subequations}
\begin{gather}
f_{1}^{\prime }=f_{1},\quad f_{j}^{\prime }=f_{j}, \\
\forall k\in \{1,\ldots ,2n\}\setminus \{1,j\},f_{k}^{\prime
}=f_{k}-g_{1j}^{-1}g_{\rule{0pt}{7.2pt}1k}f_{j}+g_{1j}^{-1}g_{\rule%
{0pt}{7.2pt}jk}f_{1}.
\end{gather}%
Since $2,j\in K_{0}$, we may rename without loss of generality $%
f_{j}^{\prime }$ as $f_{2}^{\prime }$ and $f_{2}^{\prime }$ as $%
f_{j}^{\prime }$. Let $N=M\cap \left\langle f_{1}^{\prime },f_{2}^{\prime
}\right\rangle ^{\omega }$ and let $y$ be some nonzero vector in $N$ if any: 
\end{subequations}
\begin{equation}
y=\sum_{i=3}^{r}y_{i}f_{i}^{\prime }\in N\setminus \{0\},
\end{equation}%
with $r$ the rank of $M$. Since $M$ is symplectic, there exists $z\in M$ so
that $\omega (y,z)\neq 0$:%
\begin{equation}
z=\sum_{i=1}^{r}z_{i}f_{i}^{\prime }\in M.
\end{equation}%
But with $z^{\prime }=z-z_{1}f_{1}^{\prime }-z_{2}f_{2}^{\prime }\in N$, we
also have $\omega (y,z^{\prime })=\omega (y,z)\neq 0$. Hence $y\notin
N^{\omega }$ and $N$ is symplectic. If $M$ is larger than $\left\langle
f_{1}^{\prime },f_{2}^{\prime }\right\rangle $, then $N\neq \emptyset $. We
carry out again the same reasoning until we find a free basis $h$ of $%
\mathbb{Z}
_{d}^{2n}$ the first $r$ vectors of which form a symplectic basis of $M$.
Moreover, the last $2n-r$ vectors of $h$ form a free basis $h^{\flat }$ of $%
M^{\omega }$. Up to now, we proved that $M$ is free, of even rank and such
that $M\oplus M^{\omega }=%
\mathbb{Z}
_{d}^{2n}$.

Since $\Delta (h^{\flat })=\Delta (h)$ is a unit, then in the same manner as
we showed the validity of $\mathscr{D}_{\omega }$, we see that we can apply
the entire Gram-Schmidt orthogonalisation process to $h^{\flat }$. Hence, $M$
is nearly symplectic.

Let us show the converse. Let $f$ be a symplectic basis of $%
\mathbb{Z}
_{d}^{2n}$ and $D$ the following $2n\times 2n$ diagonal matrix such that $fD$
is a basis matrix for $M$:%
\begin{equation}
D=\func{diag}(p^{s_{1}},p^{s_{2}},\ldots ,p^{s_{2n-1}},p^{s_{2n}}).
\end{equation}%
Then this other diagonal matrix $D^{\prime }$ is such that $fD^{\prime }$ is
a basis matrix for $M^{\omega }$:%
\begin{equation}
D^{\prime }=\func{diag}(p^{s-s_{2}},p^{s-s_{1}},\ldots
,p^{s-s_{2n}},p^{s-s_{2n-1}}).
\end{equation}%
Under the assumption that $M+M^{\omega }=%
\mathbb{Z}
_{d}^{2n}$, we have%
\begin{equation}
s_{1}<s\Rightarrow s-s_{1}\geq 1\Rightarrow s_{2}=0\Rightarrow s-s_{2}\geq
1\Rightarrow s_{1}=0.
\end{equation}%
The same reasoning is true starting with any $i\neq 1$ and thus $M$ is free:
Each of the $d_{ii}$'s is either $1$ or $0$. For any $i\in \{1,\ldots ,n\}$,
suppose that $f_{2i}\in M$ and let $x\in M,y\in M^{\omega }$ so that $%
f_{2i-1}=x+y$. Then%
\begin{equation}
\omega (x,f_{2i})=\omega (x+y,f_{2i})=1.
\end{equation}%
That proves that the component of $x$ along $f_{2i-1}$ is $1$ and hence $%
f_{2i-1}\in M$. By the same token, $f_{2i}$ is in $M$ if $f_{2i-1}$ is.
Therefore $M$ is symplectic and of even rank.
\end{proof}

\section*{Conclusion}

In the present work, we addressed fundamental issues about submodules over $%
\mathbb{Z}
_{d}$ motivated by the growing interest for quantum information. We saw
several kinds of reduction methods for a basis matrix of a finitely
generated submodule over $%
\mathbb{Z}
_{d}$. As a first result, we established two algorithms in order to perform
simple and symplectic reduction, namely $\mathscr{D}$ and $\mathscr{S}$\
respectively. In simple reduction, no conditions are imposed on the
computational bases, so that one is able to get a diagonal basis matrix of
the particular form specified in theorem~\ref{Simple reduction theorem}\ for
any submodule. In symplectic reduction, only symplectic computational bases
are allowed. The algorithm $\mathscr{S}$ fails to provide a diagonal basis
matrix in the general case as it meets with the rent problem. But as a
second result, this algorithm was enough for us to obtain an explicit
description of Lagrangian submodules with respect to symplectic
computational bases, as we stated in theorem~\ref{Lagrangian submodules
Reduction}. Outside its native area of study, such a description can be of
particular interest in the construction of Wigner functions over a discrete
phase space and of the corresponding marginal probabilities.

As a third result, we showed that there exist submodules with no diagonal
basis matrix with respect to any symplectic computational basis. We called
the submodules that have such a basis matrix nearly symplectic and gave an
algorithm, namely $\mathscr{D}_{\omega }$, to find a suitable symplectic
basis and the corresponding diagonal basis matrix. We also compared nearly
symplectic submodules with symplectic ones: A symplectic submodule is nearly
symplectic but its sum with its orthogonal generates $%
\mathbb{Z}
_{d}^{2n}$ as a whole. Since the core feature in the area of quantum
information we started from is the symplectic inner product, it is of
particular interest to express the relevant submodules in as simple a way as
possible whereas the way to compute a symplectic product is preserved. Thus,
we would also like to know if the tools involved in the algorithm $%
\mathscr{D}_{\omega }$ enable us to perform simultaneous reduction of
matrices for instance. Do all these patterns enable us to measure a kind of
distance between the submodules?

Let us say more about the idea behind the symplectic product. The
fundamental operation to compute such a product consists in mixing the
components of the two vectors involved following a "cross" pattern. That
basic pattern is found again in the wedge product and in its generalised
form in the computation of determinants. In a forthcoming paper, we shall
address the issue of finite projective nets over $%
\mathbb{Z}
_{d}$, where the wedge product plays a particular role in relation with
other geometrical objects pointed out by quantum information theory, as for
example MUBs. Moreover, one commonly refers to a $2\times 2$ determinant to
know whether two qubits are entangled or not. Thus we are also to see that
determinants and sums of them are a kind of measure for entanglement.

\section*{Acknowledgments}

The author has carried out this work in the frame of his PhD Thesis. He
wishes to thank Michel Planat and Metod Saniga for interesting discussions
about discrete geometry for quantum information and his advisor Maurice
Kibler for useful remarks and suggestions.

\begin{appendices}%

\section{Arithmetics in $%
\mathbb{Z}
$\ and $%
\mathbb{Z}
_{d}$}

\subsection{$\gcd $, $\func{lcm}$ and order}

In $%
\mathbb{Z}
$, the notion of greatest common divisor ($\gcd $ for short) has an
intuitive meaning. But it is equivalent to a little bit more abstract
property which will generalise to residue class rings $%
\mathbb{Z}
_{d}=%
\mathbb{Z}
/d%
\mathbb{Z}
$, $d\geq 2$. This equivalence is called B\'{e}zout's theorem. To see how it
works, note that the sets of the form $k%
\mathbb{Z}
$, $k\in 
\mathbb{Z}
$, are the sole subrings of $%
\mathbb{Z}
$. B\'{e}zout's theorem states that if $\delta $\ is the $\gcd $\ of $%
a_{1},\ldots ,a_{n}\in 
\mathbb{Z}
$:%
\begin{equation}
\delta =\bigwedge_{i=1}^{n}a_{i},
\end{equation}%
then $\delta $\ is characterised up to its sign by the set equation%
\begin{equation}
\delta 
\mathbb{Z}
=\sum_{i=1}^{n}a_{i}%
\mathbb{Z}
,  \label{gcd in Z}
\end{equation}%
that is to say $\delta 
\mathbb{Z}
$\ is the set of all linear combinations of the $a_{i}$'s over $%
\mathbb{Z}
$. We immediately deduce from that theorem Gauss's theorem for integers: If $%
a$\ divides the product $bc$\ and is coprime with $b$\ then $a$\ divides $c$%
. It is also quite obvious from B\'{e}zout's theorem that the following
three properties are equivalent:

\begin{enumerate}
\item $a$ is coprime with $d$;

\item The residue class $\overline{a}$ in the quotient ring $%
\mathbb{Z}
_{d}$ is invertible. In that case, we also say that $a$ is invertible modulo 
$d$;

\item $\overline{a}$ is a generator of $%
\mathbb{Z}
_{d}$:%
\begin{equation}
\overline{a}%
\mathbb{Z}
_{d}=\{\overline{a}x;x\in 
\mathbb{Z}
_{d}\}=%
\mathbb{Z}
_{d}.
\end{equation}
\end{enumerate}

\noindent The invertible elements of $%
\mathbb{Z}
_{d}$ are also called its units and hence their set is denoted $U(%
\mathbb{Z}
_{d})$, or $%
\mathbb{Z}
_{d}^{\ast }$.

In the case of $%
\mathbb{Z}
_{d}$, equation (\ref{gcd in Z})\ is retained in order to define a notion of 
$\gcd $.\ A residue $\overline{\delta }\in 
\mathbb{Z}
_{d}$\ is a $\gcd $ for a set of $\overline{a_{i}}$'s\ in $%
\mathbb{Z}
_{d}$\ if%
\begin{equation}
\overline{\delta }%
\mathbb{Z}
_{d}=\sum_{i=1}^{n}\overline{a_{i}}%
\mathbb{Z}
_{d}.  \label{gcd in Zd}
\end{equation}%
So, if $\delta $ is the $\gcd $ of the $a_{i}$'s, $\overline{\delta }$ is a $%
\gcd $ for the $\overline{a_{i}}$'s. As for $%
\mathbb{Z}
$, this $\gcd $\ is determined only up to an invertible mutiplier. We shall
prove that later on in Section~\ref{CRT}. The computation of a $\gcd $\ is
still associative and commutative. As is the case for $%
\mathbb{Z}
$, the $\overline{a_{i}}$'s will be said coprime if $\overline{\delta }$ is
invertible. In this case, $\overline{\delta }%
\mathbb{Z}
_{d}=%
\mathbb{Z}
_{d}$. The interpretation in terms of linear combinations is still valid.
The intuive one in terms of prime factor decomposition or division order is
also still valid if one takes into account the slight modification indicated
by the following property:%
\begin{equation}
\overline{\delta }=\bigwedge_{i=1}^{n}\overline{a_{i}}\text{ in }%
\mathbb{Z}
_{d}\text{\ iff }\delta \wedge d=\left( \bigwedge_{i=1}^{n}a_{i}\right)
\wedge d\text{ in }%
\mathbb{Z}
\text{.}  \label{gcd in Zd equivalence}
\end{equation}%
Indeed, if we come back to representatives of residue classes, definition (%
\ref{gcd in Zd}) reads%
\begin{equation}
\delta 
\mathbb{Z}
+d%
\mathbb{Z}
=\left( \sum_{i=1}^{n}a_{i}%
\mathbb{Z}
\right) +d%
\mathbb{Z}
,
\end{equation}%
which is nothing but the second member of equivalence (\ref{gcd in Zd
equivalence}). So, there is an additional $d$ in each member of that latter
expression. It means that the power $k$ of a prime factor in $\delta $ or in
any one of the $a_{i}$'s must first be replaced by the minimum of $k$ and
the power of the same prime factor in $d$. Light will be shed on that recipe
in Section~\ref{CRT} with the Chinese remainder theorem and $p$-adic
decomposition.

If $\overline{\delta }$ is a $\gcd $ for the $\overline{a_{i}}$'s, we shall
call $\overline{\delta \wedge d}$ \textit{the} $\gcd $ of the $\overline{%
a_{i}}$'s. In fact, it is a $\gcd $ and if $\overline{\delta _{1}}$ and $%
\overline{\delta _{2}}$ are two $\gcd $'s then according to (\ref{gcd in Zd
equivalence})%
\begin{equation}
\overline{\delta _{1}\wedge d}=\overline{\delta _{2}\wedge d}.
\end{equation}%
That $\gcd $ is also the first one according to the lexicographic order from 
$\overline{0}$\ to $\overline{d-1}$ since for any positive $\delta $ such
that $\overline{\delta }$ is a $\gcd $, $\delta \wedge d\leq \delta $.

In the same manner, we define a lowest common multiple ($\func{lcm}$ for
short)\ in of $a_{1},\ldots ,a_{n}\in 
\mathbb{Z}
$\ (resp.~in $\overline{a_{1}},\ldots ,\overline{a_{n}}\in 
\mathbb{Z}
_{d}$)\ to be an element $\mu _{1}$ (resp.~$\overline{\mu _{2}}$) such that%
\begin{equation}
\mu _{1}%
\mathbb{Z}
=\bigcap_{i=1}^{n}a_{i}%
\mathbb{Z}
\qquad \left( \text{resp.~}\overline{\mu _{2}}%
\mathbb{Z}
_{d}=\bigcap_{i=1}^{n}\overline{a_{i}}%
\mathbb{Z}
_{d}\right) .
\end{equation}%
The $\func{lcm}$ operation is associative and commutative in both case and
is denoted by the vee symbol $\vee $:%
\begin{equation}
\mu _{1}=\bigvee_{i=1}^{n}a_{i}\qquad \left( \text{resp.~}\overline{\mu _{2}}%
=\bigvee_{i=1}^{n}\overline{a_{i}}\right) .
\end{equation}%
Those two notions of $\func{lcm}$'s\ are related by%
\begin{equation}
\overline{\mu }=\bigvee_{i=1}^{n}\overline{a_{i}}\text{ in }%
\mathbb{Z}
_{d}\text{\ iff }\mu \wedge d=\left( \bigvee_{i=1}^{n}a_{i}\right) \wedge d%
\text{ in }%
\mathbb{Z}
\text{.}
\end{equation}%
Indeed, since the map $x\mapsto \overline{x}$ is onto, the first equality
means%
\begin{equation}
\mu 
\mathbb{Z}
+d%
\mathbb{Z}
=\bigcap_{i=1}^{n}(a_{i}%
\mathbb{Z}
+d%
\mathbb{Z}
)
\end{equation}%
and the second one means%
\begin{equation}
\mu 
\mathbb{Z}
+d%
\mathbb{Z}
=\left( \bigcap_{i=1}^{n}a_{i}%
\mathbb{Z}
\right) +d%
\mathbb{Z}
.
\end{equation}%
We are thus to prove that%
\begin{equation}
\bigcap_{i=1}^{n}(a_{i}%
\mathbb{Z}
+d%
\mathbb{Z}
)=\left( \bigcap_{i=1}^{n}a_{i}%
\mathbb{Z}
\right) +d%
\mathbb{Z}
.  \label{Distributivity of gcd and lcm}
\end{equation}%
Since all operations involved here are associative and the intersection of
two subrings is still a subring, we can prove this equality by induction. So
let us suppose that $n=2$ and let $x$\ be in the first set:%
\begin{equation}
x=k_{1}a_{1}+l_{1}d=k_{2}a_{2}+l_{2}d.
\end{equation}%
Divide each member by $a_{1}\wedge a_{2}\wedge d$:%
\begin{equation}
x^{\prime }=k_{1}a_{1}^{\prime }+l_{1}d^{\prime }=k_{2}a_{2}^{\prime
}+l_{2}d^{\prime }.
\end{equation}%
Then $a_{1}^{\prime }\wedge a_{2}^{\prime }$\ divides $k_{1}a_{1}^{\prime
}-k_{2}a_{2}^{\prime }=(l_{2}-l_{1})d^{\prime }$\ and is coprime with $%
d^{\prime }$. So there exist $n_{1},n_{2}\in 
\mathbb{Z}
$\ so that $n_{1}a_{1}^{\prime }-n_{2}a_{2}^{\prime }=l_{2}-l_{1}$. Let us
call $y=n_{1}a_{1}^{\prime }+l_{1}=n_{2}a_{2}^{\prime }+l_{2}$. We have%
\begin{equation}
x^{\prime }-yd^{\prime }=(k_{1}-n_{1}d^{\prime })a_{1}^{\prime
}=(k_{2}-n_{2}d^{\prime })a_{2}^{\prime }
\end{equation}%
and eventually%
\begin{equation}
x-yd\in \bigcap_{i=1}^{n}a_{i}%
\mathbb{Z}
.
\end{equation}%
The converse inclusion for (\ref{Distributivity of gcd and lcm}) is trivial.
Note that (\ref{Distributivity of gcd and lcm})\ was quite obvious with the
prime factor decomposition interpretation of $\gcd $\ and $\func{lcm}$\
since each of those two operations in $%
\mathbb{Z}
$ is distributive with respect to the other.

Finally, we define the order $\nu (a)$\ of $a\in 
\mathbb{Z}
_{d}$\ to be the cardinality of the subring $a%
\mathbb{Z}
_{d}=\{ka;k\in 
\mathbb{Z}
_{d}\}$. This is also the first positive natural number $n$ such that $na$
is a multiple of $d$. The only residue whose order is $1$\ is $0$, $a$ is
invertible modulo $d$ iff $\nu (a)=d$, and $\nu (a)%
\mathbb{Z}
$\ is the kernel of the linear map%
\begin{eqnarray}
\mathbb{Z}
&\longrightarrow &%
\mathbb{Z}
_{d}  \notag \\
k &\longmapsto &ka.
\end{eqnarray}%
We know from group theory that the cardinality of a subgroup $H$\ of a
finite group $G$\ is a divisor of the cardinality of $G$. For any $a\in 
\mathbb{Z}
$, since $\overline{a}%
\mathbb{Z}
_{d}$\ is a subgroup of $%
\mathbb{Z}
_{d}$, $x=d/\nu (\overline{a})$ is a well-defined integer such that the
order of $\overline{x}$ is $\nu (\overline{a})$. Let us carry out the
Euclidean division of $a$ by $x$: $a=qx+r$ with $0\leq r<x$ and suppose that 
$r\neq 0$. From the definition of $r$ and according to that latter
assumption, $\nu (\overline{r})>\nu (\overline{x})=\nu (\overline{a})$. But $%
\nu (\overline{a})\overline{r}=\nu (\overline{a})\overline{a}-\overline{q}%
(\nu (\overline{a})\overline{x})=0$ so that $\nu (\overline{r})\leq \nu (%
\overline{a})$, contradiction. Thus $\overline{a}\in \overline{x}%
\mathbb{Z}
_{d}$ and $\overline{a}%
\mathbb{Z}
_{d}\subset \overline{x}%
\mathbb{Z}
_{d}$. Since those two sets have the same cardinality they are equal and we
have just seen that no residue class $\overline{r}$ with $0\leq r<x$ can
generate this set, except for the case when $\overline{a}=\overline{x}=%
\overline{r}=0$. We deduce that $\overline{x}$ is the $\gcd $ of the
one-element family $(\overline{a})$. We shall say that it is the $\gcd $ of
the element $\overline{a}$.

So, we can compute the order of $\overline{a}$ as%
\begin{equation}
\nu (\overline{a})=\frac{d}{a\wedge d}.  \label{Calculation of order 1}
\end{equation}%
It means that if%
\begin{equation}
d=\prod_{i=1}^{n}p_{i}^{s_{i}}\text{ and }a=\prod_{i=1}^{n}p_{i}^{s_{i}^{%
\prime }}\prod_{j=1}^{m}p_{j}^{s_{j}^{\prime \prime }}  \label{PFD}
\end{equation}%
are the prime factor decompositions of $d$ and $a$, then%
\begin{equation}
\nu (\overline{a})=\prod_{i=1}^{n}p_{i}^{s_{i}-\min (s_{i},s_{i}^{\prime })}.
\label{PFD for the order}
\end{equation}%
Hence we can find again the equivalence we first deduced from B\'ezout's
theorem.

\subsection{The Chinese remainder theorem\label{CRT}}

In the previous section of this appendix, we saw that $\overline{a}%
\mathbb{Z}
_{d}=\overline{x}%
\mathbb{Z}
_{d}$ with $x=a\wedge d$. We may wonder from $\nu (\overline{a})=\nu (%
\overline{x})$ and from (\ref{PFD}) and (\ref{PFD for the order}) if there
is no invertible factor $\lambda \in 
\mathbb{Z}
_{d}$ such that $\overline{a}=\lambda \overline{x}$. Moreover, it will prove
the claim after (\ref{gcd in Zd}) that the $\gcd $ is determined up to an
invertible factor. Since if $\delta _{1}$ and $\delta _{2}$ are two possible 
$\gcd $'s, then there shall exist two invertible $\lambda _{1}$ and $\lambda
_{2}$ such that%
\begin{equation}
\delta _{k}=\lambda _{k}\overline{\left( \frac{d}{\nu (\delta _{k})}\right) }%
\text{ for }k=1,2,
\end{equation}%
and so $\delta _{2}=\lambda _{2}\lambda _{1}^{-1}\delta _{1}$. It will also
proves that for any $\gcd $ $\delta $ of the $\overline{a_{i}}$'s, $%
\overline{d/\nu (\delta )}$ is the $\gcd $ of the $\overline{a_{i}}$'s.

If for any $i$, $s_{i}^{\prime }\leq s_{i}$, the existence of $\lambda $ is
obvious: $\lambda =\overline{q}$ answers the question. But it is not any
more so obvious when there is one $i$ for which $s_{i}^{\prime }>s_{i}$. A
fundamental idea to refer to and that we use many other times in this paper
is to prove a property for $d$ a power of prime ($d=p^{s}$) and then deduce
that it is true for any composite $d$ as in (\ref{PFD}). This idea is
achieved by the so-called Chinese remainder theorem.

\begin{theorem}[Chinese remainder]
If $d=\prod_{i=1}^{n}p_{i}^{s_{i}}$ is the prime factor decomposition of $d$%
, then we have the following isomorphism of rings: 
\begin{equation}
\begin{array}{crcl}
\pi : & 
\mathbb{Z}
/d%
\mathbb{Z}
& \overset{\sim }{\longrightarrow } & \prod_{i=1}^{n}%
\mathbb{Z}
/p_{i}^{s_{i}}%
\mathbb{Z}
\\ 
& \overline{a} & \longmapsto & (a_{1},\ldots ,a_{n})%
\end{array}
\label{CRT isomorphism}
\end{equation}%
where $a_{i}=\pi _{p_{i}}(\overline{a})$ is the residue class of $a$ modulo $%
p_{i}^{s_{i}}$. Addition and multiplication on the right-hand side of (\ref%
{CRT isomorphism}) are componentwise:%
\begin{subequations}%
\begin{eqnarray}
(a_{1},\ldots ,a_{n})+(b_{1},\ldots ,b_{n}) &=&(a_{1}+b_{1},\ldots
,a_{n}+b_{n}), \\
(a_{1},\ldots ,a_{n})(b_{1},\ldots ,b_{n}) &=&(a_{1}b_{1},\ldots
,a_{n}b_{n}).  \label{CRT multiplication}
\end{eqnarray}

\end{subequations}%
\end{theorem}

The $%
\mathbb{Z}
/p_{i}^{s_{i}}%
\mathbb{Z}
$ in the theorem are called the Chinese factors of $%
\mathbb{Z}
/d%
\mathbb{Z}
$. According to (\ref{CRT multiplication}), $a$ is invertible iff all its
Chinese components $a_{i}$ are. Thus, to solve our problem, we can
equivalently search for a $\lambda _{i}$ in each Chinese factor such that $%
a_{i}=\lambda _{i}x_{i}$. Moreover, we are going to give a first cumbersome
proof of the existence of $\lambda _{i}$ to show the necessity for the $p$%
-adic decomposition in each Chinese factor. Let us suppose that $d=p^{s}$,
let $\nu =\nu (\overline{x})=\nu (\overline{a})$ and suppose that both $q$
and $q+\nu $ are noninvertible modulo $d$, that is to say $p$ divides $q$
and $q+\nu $. We are to prove this is impossible and thus there exists an
invertible $\lambda $ modulo $d$ such that $\overline{a}=\overline{\lambda }%
\overline{x}$. Indeed, since $a=qx$ and $p|q$ ($p$ divides $q$) the
properties $p^{n}|a$ and $p^{n-1}|x$ are true for $n=1$. Suppose they are
true for some positive integer $n$. We know that $\overline{x}$ is a
multiple of $\overline{a}$ in $%
\mathbb{Z}
_{d}$\ and thus there exist $k,l\in 
\mathbb{Z}
$ such that $x=ka+ld=ka+l\nu x$. Since $p|(q+\nu )-q=\nu $ and $p^{n-1}|x$, $%
p^{n}|\nu x$ and then $p^{n}|x$ due to the induction hypothethis and to the
previous expression for $x$. And since $p|q$ and $a=qx$, we deduce that $%
p^{n+1}|a$. Hence the property $p^{n}|a$ should be true for all positive
integer $n$, what is clearly nonsense when $a\neq 0$. If $a=0$, we can just
replace it by $d$. We are now going to introduce the $p$-adic decomposition
in $%
\mathbb{Z}
/p^{s}%
\mathbb{Z}
$\ and compare with a proof using it.

Let $a$ be a nonnegative integer and $p$ be prime number. Writing $a$ in
numeration basis $p$, we get the numbers $r\in 
\mathbb{N}
$ and $\alpha _{0},\ldots ,\alpha _{r}\in \{0,\ldots ,p\}$ such that 
\begin{equation}
a=\alpha _{0}+\alpha _{1}p+\cdots +\alpha _{r}p^{r}.
\end{equation}%
This is the $p$-adic decomposition of $a$. The $p$-valuation of $a$ is%
\begin{equation}
v_{p}(a)=\left\{ 
\begin{array}{l}
\min (i\in \{0,\ldots ,r\};\alpha _{i}\neq 0)\text{ for }a\neq 0, \\ 
+\infty \text{ for }a=0.%
\end{array}%
\right.
\end{equation}%
For instance, if $a=\prod_{i=1}^{n}p_{i}^{s_{i}}\neq 0$ is the prime factor
decomposition of $a$ then for any $i\in \{1,\ldots ,n\}$, $%
v_{p_{i}}(a)=s_{i} $.

Every class $\overline{a}\in 
\mathbb{Z}
/p^{s}%
\mathbb{Z}
$ is uniquely represented by an integer $a\in \{0,\ldots ,p^{s}-1\}$. So
there exist one single $(\alpha _{0},\ldots ,\alpha _{s-1})\in \{0,\ldots
,p\}^{s}$ such that%
\begin{equation}
\overline{a}=\alpha _{0}\overline{1}+\alpha _{1}\overline{p}+\cdots +\alpha
_{s-1}\overline{p}^{s-1}.
\end{equation}%
This is the $p$-adic decomposition of $\overline{a}$. The $p$-valuation of $%
\overline{a}$ is%
\begin{equation}
v_{p}(\overline{a})=\left\{ 
\begin{array}{l}
\min (i\in \{0,\ldots ,s-1\};\alpha _{i}\neq 0)\text{ for }a\neq 0, \\ 
s\text{ for }a=0.%
\end{array}%
\right.
\end{equation}%
The order of $\overline{a}$ is then $p^{s-v_{p}(\overline{a})}$ and $%
\overline{a}$ is invertible iff its valuation is $0$. Moreover, for all $%
\overline{a},\overline{b}\in 
\mathbb{Z}
/p^{s}%
\mathbb{Z}
$,%
\begin{subequations}%
\begin{eqnarray}
v_{p}(\overline{a}+\overline{b}) &\geq &\min (v_{p}(\overline{a}),v_{p}(%
\overline{b})),  \label{Valuation and sum} \\
v_{p}(\overline{a}\overline{b}) &=&\min (v_{p}(\overline{a})+v_{p}(\overline{%
b}),s),  \label{Valuation and product}
\end{eqnarray}%
\end{subequations}%
where equality in the latter formula relies on the fact that $p$ is prime.

To check their understanding of $p$-adic decomposition, the reader should be
able to see the following equalities, for any finite set $\{a_{1},\ldots
,a_{n}\}\subset 
\mathbb{Z}
$ of divisors of some $d\geq 2$:%
\begin{subequations}%
\begin{eqnarray}
\left( \bigwedge\nolimits_{i=1}^{n}a_{i}\right) \left(
\bigvee\nolimits_{i=1}^{n}d/a_{i}\right) &=&d,
\label{Calculation of order 2} \\
\left( \bigvee\nolimits_{i=1}^{n}a_{i}\right) \left(
\bigwedge\nolimits_{i=1}^{n}d/a_{i}\right) &=&d.
\end{eqnarray}

\end{subequations}%
Now, let us hark back to our search for $\lambda _{i}$. Since they are of
the same order, $a_{i}$ and $x_{i}$ are both zero or nonzero. If they are
nonzero, then according to (\ref{Valuation and product}) applied to $%
a_{i}=q_{i}x_{i}$, $q_{i}$ is of $p_{i}$-valuation $0$. Hence it is
invertible in $%
\mathbb{Z}
/p_{i}^{s_{i}}%
\mathbb{Z}
$ and we take $\lambda _{i}=q_{i}$. If they are null, then $\nu _{i}=\pi
_{p_{i}}(\overline{\nu })=\overline{1}$ and either $q_{i}$ or $q_{i}+\nu
_{i} $ is of $p_{i}$-valuation $0$ so that we get our $\lambda _{i}$. That
is a simple proof of the

\begin{lemma}
\label{Preserving order in Zd}Let $d\geq 2$ and $a,b\in 
\mathbb{Z}
_{d}$. The two following assertions are equivalent:

\begin{enumerate}
\item $a,b$ are of the same order.

\item There exist $\lambda \in U(%
\mathbb{Z}
_{d})$ such that $a=\lambda b$.
\end{enumerate}

\noindent If one of them is satified, $a$ and $b$ are said to be associated.
This is the case in particular if $a$ and $b$ are two $\gcd $'s of a same
set of elements in $%
\mathbb{Z}
_{d}$.
\end{lemma}

What about the computation of the $\gcd $ of given $a_{1},\ldots ,a_{m}\in 
\mathbb{Z}
/d%
\mathbb{Z}
$ using the Chinese remainder theorem. Let $a_{ij}=\pi _{p_{j}}(a_{i})$ for
any $i\in \{1,\ldots ,m\}$ and $j\in \{1,\ldots ,n\}$. In order to lighten
notations, we avoid the bar over residue classes in this paragraph.\ The set
to which any element belongs will be known from the context. Let also $%
\delta =\bigwedge_{i=1}^{m}a_{i}$ in $%
\mathbb{Z}
/d%
\mathbb{Z}
$ and $\delta _{j}=\pi _{p_{j}}(\delta )$. It is quite obvious that in the $%
j $-th Chinese factor of $%
\mathbb{Z}
/d%
\mathbb{Z}
$ the $\gcd $ of the $a_{ij}$'s is 
\begin{equation}
\bigwedge_{i=1}^{m}a_{ij}=p_{j}^{k_{j}},\text{ with }k_{j}=\min
(v_{p_{j}}(a_{ij});i\in \{1,\ldots ,m\})\leq s_{j}.
\end{equation}%
Indeed, if $i_{0}$ is an index for which $v_{p_{j}}(a_{i_{0}j})=k_{j}$, then 
$a_{i_{0}j}=p_{j}^{k_{j}}u$, where $u$ is invertible. Thus $p_{j}^{k_{j}}$
may be obtained as a linear combination of the $a_{ij}$'s and any linear
combination of them is a multiple of $p_{j}^{k_{j}}$. Moreover $%
p_{j}^{k_{j}}=p_{j}^{k_{j}}\wedge p_{j}^{s_{j}}$ in $%
\mathbb{Z}
$. Since $\delta $ is a linear combination of the $a_{i}$'s, $\delta _{j}$
is a linear combination of the $a_{ij}$'s and so $v_{p_{j}}(\delta _{j})\geq
k_{j}$. Then, $a_{i_{0}}$ being a multiple of $\delta $, $a_{i_{0}j}$ is a
multiple of $\delta _{j}$ and so $v_{p_{j}}(\delta _{j})=k_{j}$. Hence $%
\delta =\prod_{j=1}^{m}p_{j}^{k_{j}}$. All this is nothing but the usual way
to compute $\gcd $'s by means of prime factor decomposition.

Another useful lemma is the following one. It is not often found in
literature maybe for the crux is easy to see.

\begin{lemma}
\label{Invertible coeffs for gcd}Let $d\geq 2$ and $a,b,\delta \in 
\mathbb{Z}
_{d}$ such that $\delta $ is a $\gcd $ for $a$ and $b$. If one of the
following conditions is verified:

\begin{itemize}
\item $d$ is odd,

\item $d$ is even and $v_{2}(a)\neq v_{2}(b)$,

\item $d$ is even and $v_{2}(a)=v_{2}(b)=v_{2}(d)$;
\end{itemize}

\noindent then one can choose $u,v\in U(%
\mathbb{Z}
_{d})$ such that $\delta =ua+vb$. If not, then only $u$ or $v$ can be chosen
invertible.
\end{lemma}

\begin{proof}
In this proof, in order to distinguish classes and representatives, we shall
note $\overline{a},\overline{b},\overline{\delta }$ instead of $a,b,\delta $
as in the terms of the lemma. Using the Chinese remainder theorem, we search
for $u$ and $v$ in each Chinese factor separately. So suppose $d=p^{s}$,
with $p$ odd to begin with. Also note that owing to of lemma~\ref{Preserving
order in Zd}, it suffices to prove lemma~\ref{Invertible coeffs for gcd} for
any $\gcd $ $\overline{\delta }$ of $\overline{a}$ and $\overline{b}$. So we
will choose $\delta =a\wedge b$, taking into account the remark just
following (\ref{gcd in Zd}). By definition, there exist $u_{0},v_{0}\in 
\mathbb{Z}
$ such that $\delta =u_{0}a+v_{0}b$, and dividing by $\delta $ we obtain%
\begin{equation}
1=u_{0}a^{\prime }+v_{0}b^{\prime }  \label{Simplified Bezout relation}
\end{equation}%
where $a^{\prime }=a/\delta $, $b^{\prime }=b/\delta $. We see that $u_{0}$
and $v_{0}$ cannot be both multiples of $p$. At least one of $\overline{u_{0}%
}$ and $\overline{v_{0}}$, say $\overline{u_{0}}$, is a unit. Suppose $%
\overline{v_{0}}$ is not a unit, that is to say $v_{0}$ is a multiple of $p$%
. If $v_{0}+a^{\prime }$ were a multiple of $p$, then so would $a^{\prime }$%
, what would contradict (\ref{Simplified Bezout relation}) once more. So $%
v_{0}+a^{\prime }$ is a unit and so is $v_{0}-a^{\prime }$. Besides, if $%
u_{0}\pm b^{\prime }$ were both mutiples of $p$, so would be $2b^{\prime }$, 
$b^{\prime }$ and then $u_{0}$. We may now conclude that at least one of the
three pairs%
\begin{equation}
(\overline{u_{0}},\overline{v_{0}}),\ (\overline{u_{0}+b^{\prime }},%
\overline{v_{0}-a^{\prime }})\text{ and }(\overline{u_{0}-b^{\prime }},%
\overline{v_{0}+a^{\prime }})
\end{equation}%
is in $U(%
\mathbb{Z}
_{d})^{2}$. That proves the lemma as to the first condition.

If $p=2$ and $v_{2}(\overline{a})\neq v_{2}(\overline{b})$, then in (\ref%
{Simplified Bezout relation}), one of $a^{\prime }$ and $b^{\prime }$ is
odd, say $a^{\prime }$, and the other is even, say $b^{\prime }$. Moreover, $%
u_{0}$ has to be odd too. Then one of the two pairs%
\begin{equation}
(\overline{u_{0}},\overline{v_{0}})\text{ and }(\overline{u_{0}+b^{\prime }},%
\overline{v_{0}-a^{\prime }})
\end{equation}%
is in $U(%
\mathbb{Z}
_{d})^{2}$.

If $p=2$ and $v_{2}(\overline{a})=v_{2}(\overline{b})=v_{2}(d)$, then $%
\overline{a}=\overline{b}=\overline{\delta }=0$ and $u=v=1$ suit the lemma.

Still with $p=2$, if $v_{2}(\overline{a})=v_{2}(\overline{b})\neq v_{2}(d)$,
we have already seen that at least one of $\overline{u_{0}}$ and $\overline{%
v_{0}}$, say $\overline{u_{0}}$, is a unit. But $\overline{v_{0}}$ cannot be
a unit, since in that case $u_{0}a^{\prime }+v_{0}b^{\prime }$ should be
even. Because we only need $u_{0}a^{\prime }+v_{0}b^{\prime }$ to be odd, we
can choose which one of $\overline{u_{0}}$ and $\overline{v_{0}}$ is
invertible.
\end{proof}

By induction and associativity of $\gcd $, we have the

\begin{corollary}
\label{Invertible coeffs for gcd Cor.}Let $a_{1},a_{2},\ldots ,a_{n}\in 
\mathbb{Z}
_{d}$ and $\delta $ be one of their $\gcd $'s. For any $i\in \{1,\ldots ,n\}$%
, one can find $k_{1},k_{2},\ldots ,k_{n}\in 
\mathbb{Z}
_{d}$ with $k_{i}\in U(%
\mathbb{Z}
_{d})$\ such that%
\begin{equation}
\delta =\sum_{j=1}^{n}k_{j}a_{j}.
\end{equation}
\end{corollary}

\section{Finitely generated modules over $%
\mathbb{Z}
_{d}$\label{FGM Section}}

Let $d$ and $n$ be two positive integers with $d\geq 2$. The set product $%
\mathbb{Z}
_{d}^{n}$\ is endowed with its canonical structure of $%
\mathbb{Z}
$-module and its elements will be called vectors. Addition is componentwise:%
\begin{equation}
\begin{array}{rcl}
\mathbb{Z}
_{d}^{n}\times 
\mathbb{Z}
_{d}^{n} & \longrightarrow & 
\mathbb{Z}
_{d}^{n} \\ 
((a_{1},\ldots ,a_{n}),(b_{1},\ldots ,b_{n})) & \longmapsto & 
(a_{1}+b_{1},\ldots ,a_{n}+b_{n})%
\end{array}%
\end{equation}%
and the product map is%
\begin{equation}
\begin{array}{rcl}
\mathbb{Z}
\times 
\mathbb{Z}
_{d}^{n} & \longrightarrow & 
\mathbb{Z}
_{d}^{n} \\ 
(k,(a_{1},\ldots ,a_{n})) & \longmapsto & (ka_{1},\ldots ,ka_{n}).%
\end{array}%
\end{equation}%
This can also be denoted $k\cdot (a_{1},\ldots ,a_{n})$\ or even $%
k(a_{1},\ldots ,a_{n})$\ without a symbol. Obviously, such a product depends
only on the residue class of $k$\ modulo $d$, so that we may consider $%
\mathbb{Z}
_{d}^{n}$\ either as a $%
\mathbb{Z}
$-module\ or a $%
\mathbb{Z}
_{d}$-module. So, when the context is clear or the distinction useless, one
can avoid the bar to denote residue classes.

A submodule of $%
\mathbb{Z}
_{d}^{n}$\ is a module over $%
\mathbb{Z}
_{d}$\ included in $%
\mathbb{Z}
_{d}^{n}$. When $n=1$, submodules are called ideals of $%
\mathbb{Z}
_{d}$. Let $I$\ be a finite index set and $m=(m_{i})_{i\in I}$\ be a family
of vectors in $%
\mathbb{Z}
_{d}^{n}$. The submodule those vectors generate is the set of all their
linear combinations over $%
\mathbb{Z}
_{d}$ and is noted $\left\langle m\right\rangle $, or $\left\langle
m_{1},\ldots ,m_{r}\right\rangle $ whenever $I=\{1,\ldots ,r\}$. It is the
tiniest submodule that contains all the $m_{i}$'s. The family $m$\ is a
generating system or basis of that submodule. Moreover, any submodule of $%
\mathbb{Z}
_{d}^{n}$\ is generated by some basis, since the whole submodule itself is
such a basis. The family $m$\ is free if for all family $(c_{i})_{i\in I}$\
of elements of $%
\mathbb{Z}
_{d}$,%
\begin{equation}
\sum_{i\in I}c_{i}m_{i}=0\Longrightarrow \forall i\in I,c_{i}=0.
\end{equation}%
In other words, the linear map%
\begin{equation}
\begin{array}{crcl}
f_{m}: & 
\mathbb{Z}
_{d}^{I} & \longrightarrow & 
\mathbb{Z}
_{d}^{n} \\ 
& (c_{i})_{i\in I} & \longmapsto & \sum\nolimits_{i\in I}c_{i}m_{i}%
\end{array}%
\end{equation}%
has kernel $0$. A basis of a submodule which is also free is called a free
basis of that submodule, and a submodule for which there exists a free basis
is called a free submodule. The computational basis of $%
\mathbb{Z}
_{d}^{n}$ is of course a free basis and it will be denoted by $%
e=(e_{i})_{i=1\ldots n}$. For any vector $a$, $e_{i}^{\ast }(a)=a_{i}$ is
the $i$-th component of $a$ with respect to $e$.

A vector $a$ such that the one-element family $(a)$ is free is called a free
vector. If moreover $n=1$, then $a$ is just said regular.

A submodule $M$\ is said to be of rank $r$\ if the minimal number of vectors
needed to generate\ it is $r$. This notion of rank should not be confused
with the rank of the matrix whose columns are a set of generating vectors of 
$M$ with respect to some free basis of $%
\mathbb{Z}
_{d}^{n}$ (see \cite{Brown.93}). Those two notions of rank for submodules
and matrices do not overlap.

A minimal basis for a rank-$r$ submodule $M$\ is a basis of $M$\ with $r$\
elements. Such a basis need not be free. For instance in $%
\mathbb{Z}
_{4}^{2}$, $((2,0))$\ is a basis for the rank-$1$ submodule $\{(0,0),(2,0)\}$%
\ but is not free. But if $M$ is free, minimal and free bases are the same
ones. Indeed, let $(m_{i})_{i=1,\ldots ,r}$ be a minimal basis of $M$ and $%
(m_{i}^{\prime })_{i\in I}$ be a free basis of $M$. By minimality of $m$, $%
r\leq \left\vert I\right\vert $ and by freedom of $m^{\prime }$, $\left\vert 
\func{Im}f_{m^{\prime }}\right\vert =d^{\left\vert I\right\vert }$. So%
\begin{equation}
\left\vert M\right\vert =\left\vert \func{Im}f_{m}\right\vert \leq d^{r}\leq
d^{\left\vert I\right\vert }=\left\vert \func{Im}f_{m^{\prime }}\right\vert
=\left\vert M\right\vert .
\end{equation}%
Thus on the one hand $\left\vert \func{Im}f_{m}\right\vert =d^{r}$ and $%
f_{m} $ must be injective, so that $m$ is free. On the other hand, $%
\left\vert I\right\vert =r$ implies that $m^{\prime }$ is minimal.

Let $a=(a_{1},\ldots ,a_{n})\in 
\mathbb{Z}
_{d}^{n}$. The order $\nu (a)$\ of $a$\ is the cardinality of the set $%
\mathbb{Z}
_{d}\cdot a=\{ka;k\in 
\mathbb{Z}
_{d}\}$. The only vector whose order is $1$\ is the null vector and $a$ is a
free vector iff $\nu (a)=d$. Endly, we note that $\nu (a)%
\mathbb{Z}
$\ is the kernel of the linear map%
\begin{equation}
\begin{array}{crcl}
f: & 
\mathbb{Z}
& \longrightarrow & 
\mathbb{Z}
_{d}^{n} \\ 
& k & \longmapsto & ka.%
\end{array}%
\end{equation}%
This kernel is the intersection of the $\ker (e_{i}^{\ast }\circ f)=\nu
(a_{i})%
\mathbb{Z}
$\ and thus%
\begin{equation}
\nu (a)=\bigvee_{i=1}^{n}\nu (a_{i}).  \label{Calculation of order 3}
\end{equation}%
With (\ref{Calculation of order 1}) and (\ref{Calculation of order 2}) we
also deduce that%
\begin{equation}
\nu (a)=\frac{d}{\left( \bigwedge\nolimits_{i=1}^{n}a_{i}\right) \wedge d}.
\label{Calculation of order 4}
\end{equation}

\end{appendices}%

\bibliographystyle{unsrt}
\bibliography{LagrSubMod}

\end{document}